\documentclass[acmtog]{acmart}

\usepackage{kotex}
\usepackage{multirow}
\usepackage{tabularx}
\usepackage{subfig}
\usepackage{verbatim}
\usepackage{mathtools}
\usepackage{siunitx}
\usepackage{colortbl}
\usepackage{afterpage}
\usepackage{booktabs}
\usepackage{hyperref}

\usepackage[ruled]{algorithm2e} 

\SetAlFnt{\small}
\SetAlCapFnt{\small}
\SetAlCapNameFnt{\small}
\SetAlCapHSkip{0pt}

\setcopyright{othergov}\acmJournal{TOG}
\acmYear{2020}\acmVolume{39}\acmNumber{6}\acmArticle{222}\acmMonth{12} \acmDOI{10.1145/3414685.3417838}

\begin{document}

\title{Speech Gesture Generation from the Trimodal Context of Text, Audio, and Speaker Identity}

\author{Youngwoo Yoon}
\affiliation{\institution{ETRI, KAIST}}
\email{youngwoo@etri.re.kr}

\author{Bok Cha}
\affiliation{\institution{University of Science and Technology, ETRI}}
\email{bokc@etri.re.kr}

\author{Joo-Haeng Lee}
\affiliation{\institution{ETRI, University of Science and Technology}}
\email{joohaeng@etri.re.kr}

\author{Minsu Jang}
\affiliation{\institution{ETRI}}
\email{minsu@etri.re.kr}

\author{Jaeyeon Lee}
\affiliation{\institution{ETRI}}
\email{leejy@etri.re.kr}

\author{Jaehong Kim}
\affiliation{\institution{ETRI}}
\email{jhkim504@etri.re.kr}

\author{Geehyuk Lee}
\affiliation{\institution{KAIST}}
\email{geehyuk@gmail.com}


\begin{abstract}
For human-like agents, including virtual avatars and social robots, making proper gestures while speaking is crucial in human--agent interaction. Co-speech gestures enhance interaction experiences and make the agents look alive. However, it is difficult to generate human-like gestures due to the lack of understanding of how people gesture. Data-driven approaches attempt to learn gesticulation skills from human demonstrations, but the ambiguous and individual nature of gestures hinders learning. In this paper, we present an automatic gesture generation model that uses the multimodal context of speech text, audio, and speaker identity to reliably generate gestures. By incorporating a multimodal context and an adversarial training scheme, the proposed model outputs gestures that are human-like and that match with speech content and rhythm. We also introduce a new quantitative evaluation metric for gesture generation models. Experiments with the introduced metric and subjective human evaluation showed that the proposed gesture generation model is better than existing end-to-end generation models. We further confirm that our model is able to work with synthesized audio in a scenario where contexts are constrained, and show that different gesture styles can be generated for the same speech by specifying different speaker identities in the style embedding space that is learned from videos of various speakers. All the code and data is available at \url{https://github.com/ai4r/Gesture-Generation-from-Trimodal-Context}.
\end{abstract}

%
%

%
%


\begin{teaserfigure}
  \centering
  \includegraphics[width=\textwidth]{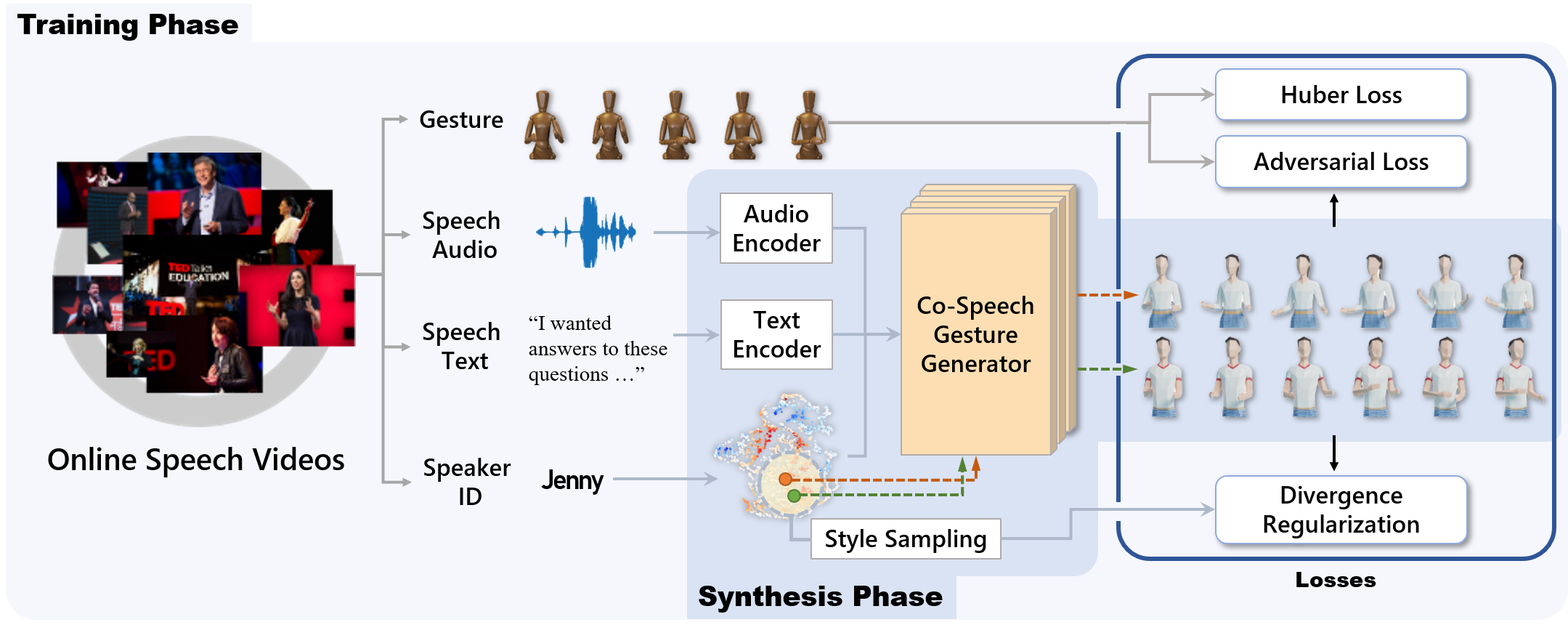}
  \caption{Overview of the proposed gesture generation model that considers the trimodality of speech text, audio, and speaker identity. The model is trained on online speech videos demonstrating co-speech gestures. At the synthesis phase, we can manipulate gesture styles by sampling a style vector from the learned style embedding space.}
  \label{fig:teaser}
\end{teaserfigure}
\maketitle
\section{Introduction}

The continued development of graphics and robotics technology has prompted the development of artificial embodied agents, such as virtual avatars and social robots, as a popular interaction medium. One of the merits of the embodied agent is its nonverbal behavior, including facial expressions, hand gestures, and body gestures. In the present paper, we focus on upper-body gestures that occur with speech. Such co-speech gestures are a representative example of nonverbal communication between people. Appropriate use of gestures is helpful for understanding speech \cite{mcneill1992hand} and increases persuasion and credibility \cite{burgoon1990nonverbal}. Gestures are important not only in human--human interaction, but also in human--machine interaction. Gestures performed by artificial agents help a listener to concentrate and understand utterances \cite{bremner2011effects} and improve the intimacy between humans and agents \cite{wilson2017hand}.

Interactive artificial agents, such as game characters, virtual avatars, and social robots, need to generate gestures in real time in accord with their speech. Automatically generating co-speech gestures is a difficult problem because machines must be able to understand speech, gestures, and the relationship between them. Two representative gesture generation methods are rule-based and data-driven approaches \cite{kopp2006towards, kipp2005gesture}. The rule-based approach, as the name suggests, defines various rules mapping speech to gestures; it requires considerable human effort to define the rules, but it is widely used in commercial robots because these models are relatively simple and intuitive. The data-driven approach learns gesticulation skills from human demonstrations. This approach requires more complex models and large amounts of data, but they do not require human effort in designing rules. As large gesture datasets are becoming more available, research on data-driven approaches is increasing, e.g., \cite{kipp2005gesture, huang2014learning, chiu2015predicting, ginosar2019gestures, yoon2019robots}.

One data-driven approach, called the end-to-end method \cite{ginosar2019gestures, yoon2019robots}, is unlike others in that it uses raw gesture data without intermediate representation such as predefined unit gestures. Such less restrictive representation increases the method’s expressive capacity, enabling it to generate more natural gestures. Previous studies have successfully demonstrated end-to-end gesture generation methods. However, they were limited by their consideration of only a single modality, either speech audio or text. Since human gestures are associated with various factors, such as speech content, speech audio, interlocutor interaction, individual personality, and surrounding environment, generating gestures from a single speech modality can produce a very limited model. In the study of human gestures \cite{mcneill1992hand}, researchers have defined four categories, called iconic, metaphoric, deictic, and beat gestures, which are related to different contexts. Iconic gestures illustrate physical actions or properties (e.g., raising one’s hands while saying ``tall'') and metaphoric gestures describes abstract concepts (e.g., moving one’s hands up and down to depict a wall while saying ``constraint''). Both iconic and metaphoric gestures are highly related to the speech lexicon. Deictic gestures are indicative motions that point to a specific target or space, and are related to both the speech lexicon and the spatial context in which the gesture is made. Beat gestures are rhythmic movements that are closely related to the speech audio. In addition, even with the same speech and in the same surrounding environment, each person makes different gestures every time due to inter-person and intra-person variability of human gestures, and the inter-person variability may be attributed to individual personality. Various modalities related to speech should be considered in order to generate more meaningful and human-like gestures. 


In the present study, we propose an end-to-end gesture generation model that uses the multimodal context of text for speech content, audio for speech rhythm, and speaker identity (ID) for style variations. To integrate these multiple modalities, a temporally synchronized encoder--decoder architecture is devised based on the property of temporal synchrony found between speech and gestures in human gesture studies \cite{mcneill2008gesture, chu2014synchronization}. We experimentally confirm that each modality is effective. Especially, a style embedding space is learned from speaker IDs to reflect inter-person variability, so we can create different styles of gestures for the same speech by sampling different points in the style embedding space. Figure \ref{fig:teaser} provides an overview of the proposed gesture generation model and its training. The model is trained on a dataset derived from online videos exhibiting speech gestures with a training objective to generate human-like and diverse gestures. Our task is to develop a general gesture generator, a model that is supposed to generate convincing gestures for previously unseen speech.

A major hurdle in gesture generation studies is determining how to evaluate results. There is no single ground truth in gesture generation and well-defined evaluation methods are not yet available. Subjective human evaluation is the most reasonable method, but it is not cost effective and difficult to reproduce results. Some studies have used the mean absolute error (MAE) of the positions of body joints between human gesture examples and generated gestures for the same speech \cite{ginosar2019gestures, joo2019towards}. The MAE evaluation method is objective and reproducible, though it is hard to ascertain to what extent the MAE between joints correlates with perceived gesture quality. In the present paper, we apply the Fr\'{e}chet inception distance (FID) concept proposed in image generation research \cite{heusel2017gans} to our problem of gesture generation. FID compares fitted distributions on a latent image feature space between the sets of real and generated images. We introduce the Fr\'{e}chet gesture distance (FGD), which compares samples on a latent gesture feature space. With synthetic noisy data and comparing to human judgements, we validate that the proposed metrics are more perceptually plausible than computing the MAE between gestures. 

Our contributions can be summarized as follows:
\begin{itemize}
  \item A new gesture generation model using a trimodal context of speech text, audio, and speaker identity. To the best of our knowledge, this is the first end-to-end approach using trimodality to generate co-speech gestures.
  \item The proposal and validation of a new objective evaluation metric for gesture generation models.
  \item Extensive experiments to verify the usability of the proposed model. We show style manipulations with the trained style embedding space, the model’s response to altered speech text, and the gestures’ incorporation with synthesized audio.
\end{itemize}

The remainder of this paper is organized as follows. We first introduce related research (Section \ref{sec:relatedworks}), then describe the proposed model (Section \ref{sec:method}) and its training in detail (Section \ref{sec:training}). Section \ref{sec:metric} introduces a metric for evaluating gesture generative models and Section \ref{sec:human_evaluation} describes human evaluation to validate the proposed metric. Section \ref{sec:experiment} presents qualitative and quantitative results. Finally, Section \ref{sec:discussion} concludes the paper with a discussion of the limitations and future direction of the present research.

\section{Related Work} \label{sec:relatedworks}

We first review automatic co-speech gesture generation methods for artificial agents. Next, we introduce previous data-driven gesture generation approaches. Related work discussing gesture styles, multimodality, and evaluation methods are also introduced.

\paragraph{Co-speech Gesture Generation for Artificial Agents}

Motion capture and retargeting human motions to artificial agents is widely used to generate motions, especially in commercial systems, because of its high-quality motion from human actors \cite{menache2000understanding}. Nonverbal behavior can also be generated by retargeting human motion \cite{kim2019c}. However, the motion capture method has a critical limitation: the motion should be recorded beforehand. Therefore, the motion capture method can only be used in movies or games that have specified scripts. Interactive applications, in which the agents interact with humans with various speech utterances in real time, mostly use automatic gesture generation methods. The typical automatic generation method is rule-based generation \cite{cassell2004beat, kopp2006towards, marsella2013virtual}. For example, the robots NAO and Pepper \cite{naoqi} have a predefined set of unit gestures and have rules that connect speech words and unit gestures. This rule-based method requires human effort to design the unit gestures and hundreds of mapping rules. Research into data-driven methods has aimed to reduce the human effort required for rule generation; these methods find gesture generation rules in data using machine learning techniques. Probabilistic modeling for speech--gesture mapping has also been studied \cite{kipp2005gesture, levine2010gesture, huang2014learning} and a neural classification model selecting a proper gesture for given speech context \cite{chiu2015predicting} was also proposed. The review paper \cite{wagner2014gesture} provides a comprehensive summary of the gesture generation research and rule-based approaches.

\paragraph{End-to-end Gesture Generation Methods}

Gesture generation is a complex problem that requires understanding speech, gestures, and their relationships. To reduce the complexity of this task, previous data-driven models have divided speech into discrete topics \cite{sadoughi2019speech} or represented gestures as predefined unit gestures \cite{kipp2005gesture, levine2010gesture, huang2014learning}. However, with recent advancements in deep learning, an end-to-end approach using raw gesture data is possible. There are studies using the end-to-end approach \cite{ginosar2019gestures, yoon2019robots, ferstl2019multi, kucherenko2019analyzing, kucherenko2020gesticulator} that have formulated gesture generation as a regression problem rather than a classification problem. This continuous gesture generation does not require crafting unit gestures and their rules and also removes the restriction that gesture expressions must be selected from predetermined unit gestures. 

One study used an attentional Seq2Seq network that generates a sequence of upper body poses from speech text \cite{yoon2019robots}. The network consists of a text encoder that processes speech text and a gesture decoder that generates a pose sequence. Other studies generated gestures from speech audio \cite{ginosar2019gestures, ferstl2019multi, kucherenko2019analyzing}. These audio-based generators also based on the neural architectures generating a sequence of poses, and some studies used adversarial loss to guide generated gestures to become similar to actual human gestures. The main difference between the previous models is the use of different speech modalities. Both semantics and acoustics are important for generating co-speech gestures \cite{mcneill1992hand}, so, in this paper, we propose a model that uses multimodal speech information, audio and text together. Note that there is a concurrent work considering both audio and text information, but it trained and validated the generative model on a limited dataset of a single actor \cite{kucherenko2020gesticulator}. 

\paragraph{Learning Styles of Gestures}
People make different gestures even when they say the same words \cite{hostetter2012effects}. Similarly, artificial agents must also learn different styles of gestures. The agents should be able to make extrovert- or introvert-style gestures according to their emotional states, interaction history, user preferences, and other factors. Stylized gestures also give the agents a unique identity similar to appearances and voices. Previous studies have attempted to generate such stylized gestures \cite{neff2008gesture, levine2010gesture, ginosar2019gestures}. In these studies, generative models were trained separately for each speaker or style. This approach is an obvious way of learning individual styles, but requires a substantial amount of training data for each individual style. Because of this limitation, only three and ten individual styles were trained in \cite{levine2010gesture} and \cite{ginosar2019gestures}, respectively. In the present study, we aim to build a style embedding space, so that we can manipulate styles through sampling the space into which different styles are embedded, rather than replicating a particular style as the previous papers did. Another study proposed more detailed style manipulation by using control signals of hand position, motion speed, or moving space \cite{alexanderson2020style}.

\paragraph{Processing Multimodal Data}
The present study considers four modalities: text, audio, gesture motion, and speaker identity. Generally, multimodal data processing includes the representation of each modality, alignment between modalities, and translation between modalities \cite{baltruvsaitis2018multimodal}. There are two approaches to representation: one is that all modalities share the same representation and the other is that modalities are represented separately, and later alignment or translation stages integrate them. We can find both representation approaches related to gesture generation. A study by \cite{ahuja2019language2pose} represented both human motion and descriptive text as vectors in the same embedding space. In other studies, different representations are used for different modalities \cite{roddy2018multimodal, sadoughi2019speech}. We use separate representations, owing to the difficulty of learning a cross-modal representation for co-speech gestures arising from the weak and ambiguous relationship between speech and gestures.

Alignment between modalities is also an important factor for time-series data. In \cite{ginosar2019gestures}, a feature vector encoding input speech was passed to a decoder to generate gestures, and the alignment between the modalities is not explicitly handled. A neural encoder and decoder implicitly processed the alignment as well as the translation from speech to gesture. In \cite{yoon2019robots}, a similar encoder--decoder architecture was used, but they guided the model to learn sequential alignment more explicitly by incorporating an attention mechanism \cite{bahdanau2014neural}. In \cite{kucherenko2020gesticulator}, speech audio and text were aligned but not with gestures. Our model uses explicitly aligned speech and gesture because speech and gesture are synchronized temporally \cite{chu2014synchronization}, allowing the network to concentrate on the translation from input speech to gestures.

\paragraph{Evaluating Generative Models}
Recently, as research into generative models has expanded, interest in evaluating generative models has increased. In a generation problem considering speech synthesis, image generation, and conversational text generation, human evaluation is the most plausible evaluation method because there is no clear ground truth to compare with. However, the results of human evaluation cannot easily be reproduced. A reliable computational evaluation metric is necessary for reproducible comparisons with state-of-the-art models and would accelerate research. Previous studies have measured gesture differences between generated and human gestures \cite{ginosar2019gestures, joo2019towards}, though this method is limited because pose-level differences do not measure the perceptual quality of the generated gestures. Some studies have used other metrics to evaluate human motion, for example, the motion statistics of jerk and acceleration \cite{kucherenko2019analyzing} and Laban parameters from a study of choreography \cite{aristidou2015folk}. However, the aforementioned metrics compute distances for each sample, so they cannot measure how the generated results are diversified, which is crucial in generation problems. In the image generation problem, the inception score \cite{salimans2016improved} and FID \cite{heusel2017gans} have recently become de facto evaluation metrics because they can measure the diversity of generated samples as well as their quality, and this concept was successfully applied to other generation problems \cite{unterthiner2018towards, kilgour2018fr}. In this study, we have applied the concept of FID to the gesture generation problem to measure both perceptual quality and diversity. 

\section{Method} \label{sec:method}

\subsection{Overall Architecture}

\begin{figure}
  \centering
  \includegraphics[width=\linewidth]{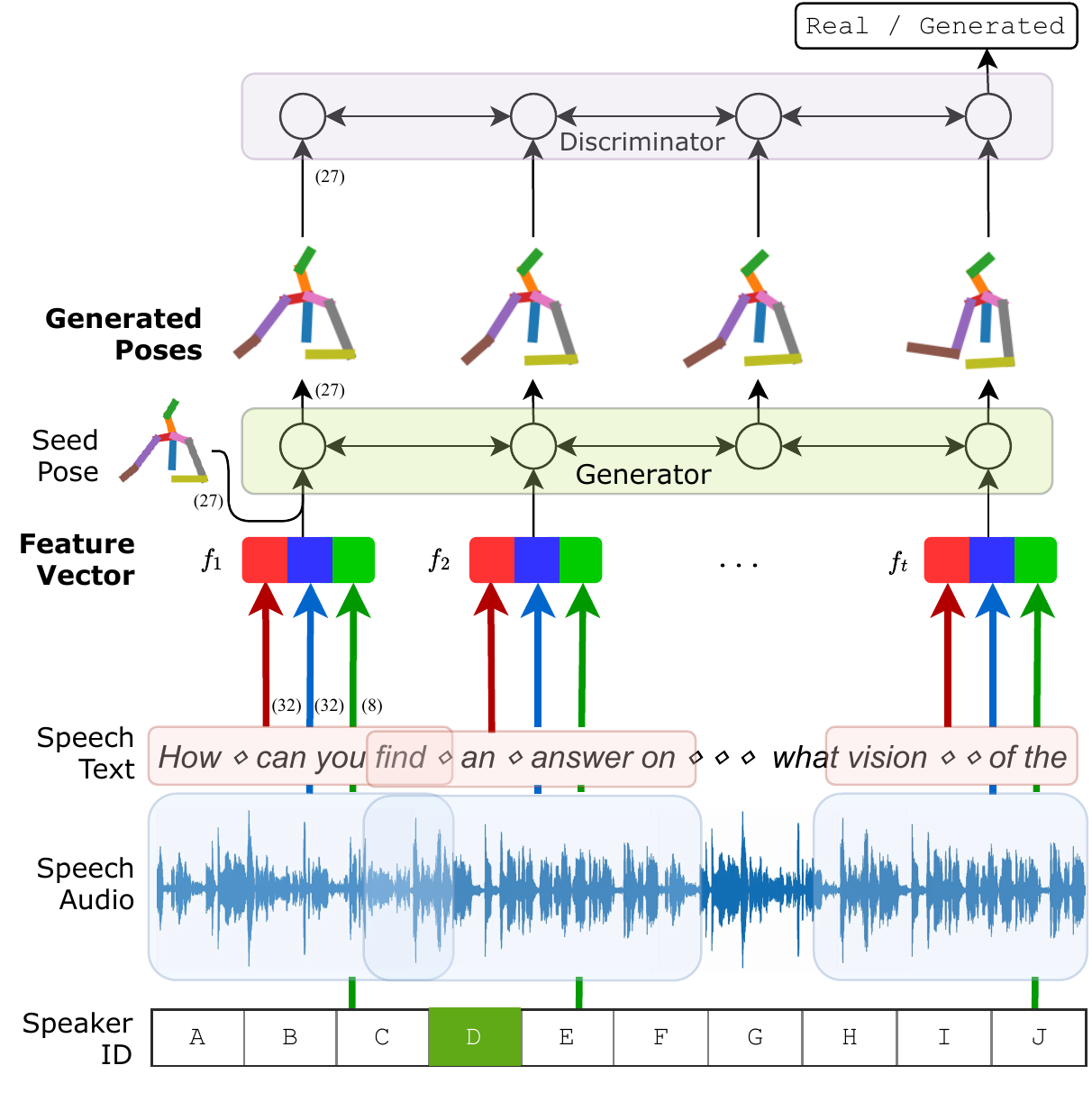}
  \caption{The architecture of the proposed gesture generation model. The generator generates a sequence of human poses from a sequence of context feature vectors that contain the encoded features of speech text, speech audio, and speaker identity (ID). The features of text, audio, and speaker ID are depicted as red, blue, and green arrows, respectively. The seed poses are also used to ensure continuity between consecutive syntheses. The discriminator is a binary classifier that distinguishes between real human gestures and generated gestures. The number in parentheses indicates the data dimension. The poses are in 27 dimensions since there are nine directional vectors in 3D coordinates.}
  \label{fig:architecture}
\end{figure}

Gesture generation in this paper is a translation problem that generates co-speech gestures from a given speech context. Our goal is to generate gestures that are human-like and match well with any given speech. We propose a neural network architecture consisting of three encoders for input speech modalities and a decoder for gesture generation. Figure \ref{fig:architecture} shows the overall architecture. Three modalities---text, audio, and speaker identity (ID)---are encoded with different encoder networks and transferred to the gesture generator.

A gesture is represented as a sequence of human poses, and the generator, which is a recurrent neural network, generates poses frame-by-frame from an input sequence of feature vectors containing encoded speech context. Speech and gestures are temporally synchronized \cite{mcneill2008gesture, chu2014synchronization}, so we configured the generator to use part of the speech text and audio near the current time step instead of the whole speech context. Gesture style does not change in the short term, so the same speaker ID is used throughout the synthesis. In addition, we used seed poses for the first few frames for better continuity between consecutive syntheses. See appendix A for the figures of detailed architecture.

\subsection{Encoding Speech Context}
This section describes how the speech modalities of text, audio, and speaker ID are represented and the details of the encoder networks. We have four modalities, including the output gesture, in different time resolution. We first ensure that all input data have the same time resolution as the output gestures, so all modalities share the same time steps and the proposed sequential model (Figure \ref{fig:architecture}) can process speech input and generate poses frame by frame. 

The speech text is a word sequence, with the number of words varying according to speech speed. We insert padding tokens ($\diamond$) into the word sequence to make a padded word sequence (${word}_1$, ${word}_2$, ..., ${word}_t$) that is the same length as the gestures. Here, $t$ is the number of poses in a synthesis (fixed as 34 throughout the paper, see Section \ref{sec:training}). We assume the exact utterance time of words is known, so the padding token is inserted to make the words temporally match the gestures. For instance, for the speech text \textit{``I love you''}, if there were a short pause between \textit{``I''} and \textit{``love''}, then the padded word sequence would be \textit{``I $\diamond$ $\diamond$ love you''} when $t$ is 5. All words in the padded word sequence are then transformed into word vectors in 300 dimensions via a word embedding layer. Next, these word vectors are encoded by a temporal convolutional network (TCN) \cite{BaiTCN2018} to make 32-D feature vectors for speech text modality ($f_1^\textrm{text}$, $f_2^\textrm{text}$, ..., $f_t^\textrm{text}$). TCN processes sequential data through convolutional operations, and showed competitive results over the recurrent neural networks in diverse problems \cite{BaiTCN2018}. In this paper, we used a four-layered TCN, where each $f^\textrm{text}$ has a receptive field of 16. Thus, $f_i^\textrm{text}$ encodes 16 padded words around at time step $i$. For our training dataset the average and the largest number of non-padding words in this receptive field were 3.9 and 16, respectively. 

We used FastText \cite{bojanowski2016enriching}, a pretrained word embedding, and update these embeddings during training. There was the concern that word embeddings pretrained by filling a missing word in a sentence \cite{mikolov2013distributed} may not suitable to gesture generation. For instance, if we query words that are close to \textit{large}, then \textit{small} appears in the top-3 list in both GloVe \cite{pennington2014glove} and FastText \cite{bojanowski2016enriching} even though they have opposite meanings. This problem with pretrained word embedding has also been raised in text-based sentiment analysis, where the sentiment of words is important \cite{fu2018learning}. We tested three different settings: 1) pretrained embeddings without weight updating, 2) pretrained embeddings with fine-tuning weights, and 3) learning word embeddings from scratch. In our problem, using pretrained embeddings with fine-tuning was the most successful. FastText \cite{bojanowski2016enriching} was favored over GloVe \cite{pennington2014glove} since FastText is using subword information so that it gives accurate representation for unseen words.

For the speech audio modality, a raw audio waveform goes through cascaded one-dimensional (1D) convolutional layers to generate a sequence of 32-D feature vectors ($f_1^\textrm{audio}$, $f_2^\textrm{audio}$, ..., $f_t^\textrm{audio}$). Audio frequency is usually fixed, so we adjusted the sizes, strides, and padding in the convolutional layers to obtain equally many audio feature vectors as there were output motion frames. In our experiments, each feature vector had a receptive field of about a quarter of a second. The quarter-second receptive field may not be large enough to cover occasional asynchrony between speech and gesture (the standard deviation of the temporal
differences is about a half second according to \cite{bergmann2011relation}), but our use of a bidirectional GRU in the gesture generator that sends information forwards and backwards can compensate for the asynchrony.

The model also uses speaker IDs to learn a style embedding space. Human gestures are not the same even for the same speech. We utilize the speaker IDs to reflect characteristics of each speaker in the dataset, and we call this individuality as `style' in the present paper. Note that our purpose is to build an embedding space capturing different styles not to replicate gestures of each speaker. The speaker IDs are represented as one-hot vectors where only one element of a selected speaker is nonzero. A set of fully connected layers maps a speaker ID to a style embedding space of much smaller dimension (8 in the present study). To make the style embedding space more interpretable, variational inference \cite{kingma2013auto, rezende2014stochastic} that uses a probabilistic sampling process is used. The same feature vector $f^\textrm{style}$ on the style embedding space is used for all time steps in a synthesis.

\subsection{Gesture Generator}
The generator $G(\cdot)$ takes encoded features as input and generates gestures. The gesture is a sequence of human poses $p_i$ consisting of 10 upper body joints (spine, head, nose, neck, L/R shoulders, L/R elbows, and L/R wrists). All poses were spine-centered. When we train the model, we represent each pose as directional vectors which represent the relative positions of the child joints from the parent joints. There are nine directional vectors for spine--neck, neck--nose, nose--head, neck--R/L shoulders, R/L shoulders--R/L elbows, and R/L elbows--R/L wrists. The directional vectors are favored for training the proposed model because this representation is less affected by bone lengths and root motion. In the representation of joint coordinates, a small translation of neck, which is the parent joint of both arms, can have an excessive effect on all coordinates of the arms. We denote human poses represented as directional vectors by $d_i$, and all directional vectors were normalized to the unit length. We note that forearm twists were not considered in this paper.

For gesture generation, we use a multilayered bidirectional gated recurrent unit (GRU) network \cite{cho2014learning}. Encoded features of speech text, audio, and speaker ID are concatenated to form a concatenated feature vector $f_i$ = ($f_i^\textrm{text}$, $f_i^\textrm{audio}$, $f^\textrm{style}$) for each time instant $i$. The generator takes the feature vector $f_i$ as input and generates the next pose $\hat{d}_{i+1}$ iteratively. 

For a long speech, the speech is divided into 2-second chunks and the generator synthesizes gestures for each chunk. The use of seed poses helps to make transitions between consecutive syntheses smooth. Seed poses $d_{i=1,...,4}$, the last four frames of the previous synthesis, are concatenated with the feature vector for the early four frames of the next synthesis as $(f_i, d_i)$, and an additional bit is used to indicate the presence of a seed pose.

\subsection{Adversarial Scheme}
An adversarial scheme \cite{goodfellow2014generative} is applied in training the model to generate more realistic gestures. The adversarial scheme uses a discriminator, which is a binary classifier distinguishing between real and generated gestures. By alternate optimization of generator and discriminator, the generator improves its performance to fool the discriminator. For the discriminator, we use a multilayered bidirectional GRU that outputs binary output for each time step. A fully connected layer aggregate the $t$ binary outputs and gives a final binary (real or generated gesture) decision.

\section{Training with ``in-the-wild'' Videos} \label{sec:training}

\subsection{TED Gesture Dataset}
The gesture generation model is trained on the TED gesture dataset \cite{yoon2019robots}, which is a large-scale, English-language dataset for data-driven gesture generation research. The dataset includes speech from various speakers, so it is suitable for learning individual gesture styles. We added 471 additional TED videos to the data of \cite{yoon2019robots}, for a total of 1,766 videos. Extracted human poses from TED videos, speech audio, and transcribed English speech text are available. We further converted all human poses to 3D by using the 3D pose estimator \cite{pavllo20193d} which convert a sequence of 2D poses into 3D poses. The pose estimator uses temporal convolutions that lead to temporally coherent results despite of a few of inaccurate 2D poses. We used the manual speech transcriptions available on each TED talk, with onset timestamps of each word extracted using the Gentle forced aligner \cite{ochshorn2016gentle} to insert padding tokens. The forced aligner reported successful alignment of 97\% of the total words.

From the videos, only the sections of videos in which upper body gestures were clearly visible were extracted; the total duration of the valid data was 97 h. The gesture poses were resampled at 15 frames per second, and each training sample having 34 frames was sampled with a stride of 10 from the valid video sections. The initial four frames were used as seed poses and the model was trained to generate the remaining 30 poses (2 seconds). We excluded non-informative samples having little motion (i.e., low variance of a sequence of poses) and erratic samples having lying poses (i.e., low angle of the spine--neck vector).

The dataset was divided into training, validation, and test sets. The division was done at the video level. Because all presentations in the TED dataset were given by different speakers, the number of unique speaker IDs is the same as the number of videos and there is no overlap of speaker IDs between split sets. We used the training set for training the model, the validation set for tuning the systems, and the test set for qualitative results and human evaluation. The final number of 34-frame sequences in each data partition were 199,384; 26,795; and 25,930. 

\subsection{Training Loss Function}
The model is trained using the losses below. We use $L_{G}$ to train the encoders and gesture generator and $L_{D}$ to train the discriminator.

\begin{equation}
\label{eqn:loss}
L_{G}=\alpha \cdot L_G^\textrm{Huber}  + \beta \cdot L_G^\textrm{NSGAN} + \gamma \cdot L_G^\textrm{style} + \lambda \cdot L_G^\textrm{KLD}
\end{equation}

\begin{equation}
\label{eqn:loss_l1}
L_G^\textrm{Huber} = \mathbb{E}[{\frac{1}{t}}\sum_{i=1}^{t}\textrm{HuberLoss}(d_{i}, \hat{d}_{i})]
\end{equation}

\begin{equation}
\label{eqn:loss_G}
L_G^\textrm{NSGAN} = -\mathbb{E}[\log(D(\hat{d}))]
\end{equation}

\begin{equation}
\label{eqn:loss_vid_reg}
L_G^\textrm{style} = -\mathbb{E} \left[ \textrm{min} \left( \dfrac{\splitdfrac{
\textrm{HuberLoss}(G(f^\textrm{text}, f^\textrm{audio}, f^\mathrm{style_1})}{\quad-G(f^\textrm{text}, f^\textrm{audio}, f^\mathrm{style_2}))
}}{\lVert f^\mathrm{style_1}-f^\mathrm{style_2} \rVert_1} , \tau \right) \right]
\end{equation}

\begin{equation}
\label{eqn:loss_D}
L_D = -\mathbb{E}[\log(D(d))]-\mathbb{E}[\log(1-D(\hat{d}))]
\end{equation}

\vspace{3mm}
\noindent where $t$ is the length of the gesture sequence, $d_{i}$ represents the $i$th pose, represented as directional vectors, in a training sample. When training the encoder and gesture generator, we minimized the difference between human poses $d$ in the training examples and the corresponding generated poses $\hat{d}$ using the Huber loss \cite{huber1964robust}. This loss $L_G^\textrm{Huber}$ can be interpreted as a once-differentiable combination of the L1 and L2 losses, and is therefore sometimes called the smooth L1 loss. The adversarial losses $L_G^\textrm{NSGAN}$ and $L_D$ are from the non-saturating generative adversarial network (NS-GAN) \cite{goodfellow2014generative}. We use sample mean to approximate the expectation terms.

A generative model conditioned on multiple input contexts often suffers from posterior collapse where weak context is ignored. In the proposed model, various gestures can be generated only from text and audio, so the style features from speaker IDs might be ignored during training. Thus, we use diversity regularization \cite{yang2019diversity} to avoid ignoring style features. $L_G^\textrm{style}$ is the Huber loss between the gestures generated from different style features normalized by the differences of the two style features, so it guides style features in the embedding space to generate different gestures. $\tau$ is for value clamping for numerical stability. In Equation \ref{eqn:loss_vid_reg}, $f^\mathrm{style_1}$ is the style feature corresponding to the speaker ID of a training sample, and $f^\mathrm{style_2}$ is the style feature for a speaker ID selected randomly. $L_G^\textrm{KLD}$, the Kullback–Leibler (KL) divergence between $\mathcal{N}(0, I)$ and the style embedding space assumed Gaussian, prevents the style embedding space from being too sparse \cite{kingma2013auto}.

$L_{D}$ is to train the discriminator $D$, and the generator and discriminator are alternately updated with $L_{G}$ and $L_{D}$ as in conventional GAN training \cite{goodfellow2014generative}. $D(\cdot)$ is trained to output 1 for human gestures and 0 for generated gestures.

The model was trained for 100 epochs. An Adam optimizer with $\beta_1=0.5$ and $\beta_2=0.999$ was used, and the learning rate was 0.0005. Weights for the loss terms were determined experimentally ($\alpha=500$, $\beta=5$, $\gamma=0.05$, and $\lambda=0.1$). In addition, there was a warm-up period of 10 epochs in which the adversarial loss was not used ($\beta=0$). $\tau$ was 1000.

The trained encoders and generator are used at the synthesis stage. As the model is lightweight enough, the synthesis can be done in real time. A single synthesis generating 30 poses takes 10 ms on a GPU (NVIDIA RTX 2080 Ti) and 80 ms on a CPU (Intel i7-5930K).
\section{Objective Evaluation Metric} \label{sec:metric}
It is difficult to evaluate gesture generation models objectively because no perceptual quality metric is available for human gestures. Although a human evaluation method in which participants rate generated gestures subjectively is possible, objective evaluation metrics are still required for fair and reproducible comparisons between state-of-the-art models. No proper and widely used evaluation metric is yet available for the gesture generation problem. 

Image generation studies have proposed the FID metric \cite{heusel2017gans}. Latent image features are extracted from the generated images using a pretrained feature extractor and FID calculates the Fr\'{e}chet distance between the distributions of the features of real and generated images. Because FID uses feature vectors that describe visual characteristics well, FID is more perceptually appropriate than measurements over raw pixel spaces. FID can also measure the diversity of the generated samples by using the samples’ distribution rather than simply averaging the differences between the real and generated samples. The diversity of generation has been thought to be one of major factors in evaluating generative models \cite{borji2019pros}. Diversity is also crucial for the gesture generation problem because the use of repetitive gestures makes artificial agents look dull.

\subsection{Fr\'{e}chet Gesture Distance}
In applying the concept of FID to the gesture generation problem, there is a hurdle that no general feature extractor is available for gesture data. The paper proposing FID used an inception network trained on the ImageNet database for image classification, but there is no analog of the pretrained inception network for gesture motion data to the best of our knowledge. Accordingly, we trained a feature extractor based on autoencoding \cite{rumelhart1985learning}, which can be trained in unsupervised manner. The feature extractor consists of a convolutional encoder and decoder; the encoder encodes a sequence of direction vectors $d$ to a latent feature $z^{gesture}$ and the decoder then attempts to restore the original pose sequence from the latent $z^{gesture}$ (see appendix A for the detailed architecture). This unsupervised learning is unlike the supervised learning of the inception network used in FID. However, both supervised and unsupervised learning have proven to be effective for learning perceptual quality metrics \cite{zhang2018unreasonable}. 

The encoder part of the trained autoencoder was used as a feature extractor. We defined FGD($X, \hat{X}$) as the Fr\'{e}chet distance between the Gaussian mean and covariance of the latent features of human gestures $X$ and the Gaussian mean and covariance of the latent features of the generated gestures $\hat{X}$ as follows:

\begin{equation} 
\textrm{FGD}(X, \hat{X})=\lVert \mu_r - \mu_g \rVert^2 + \textrm{Tr}(\Sigma_r + \Sigma_g - 2(\Sigma_r \Sigma_g)^{1/2})
\end{equation}

\noindent where $\mu_r$ and $\Sigma_r$ are the first and second moments of the latent feature distribution $Z_r$ of real human gestures $X$, and $\mu_g$ and $\Sigma_g$ are the first and second moment of the latent feature distribution $Z_g$ of generated gestures $\hat{X}$.

For training the feature extractor, we used the Human3.6M dataset \cite{ionescu2013human3} containing motion capture data of 7 different actors and 17 different scenarios including discussion and making purchases showing co-speech gestures. The total duration of the training data was about 175 m. All poses were frontalized based on two hip joints. 

\subsection{Experiment with Synthetic Noisy Data} \label{sec:validation}

\def\imagetop#1{$\vcenter{\null\hbox{#1}}$}
\begin{figure}
    \begin{tabular}{l l} 
    \imagetop{(a)} & \imagetop{\includegraphics[trim=200 45 280 50 ,clip,width=0.87\linewidth]{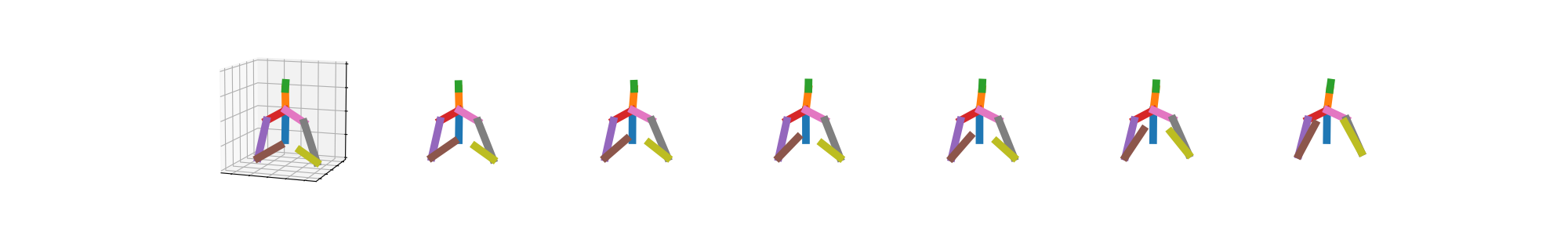}}\\
    \imagetop{(b)} & \imagetop{\includegraphics[trim=200 50 280 50 ,clip,width=0.87\linewidth]{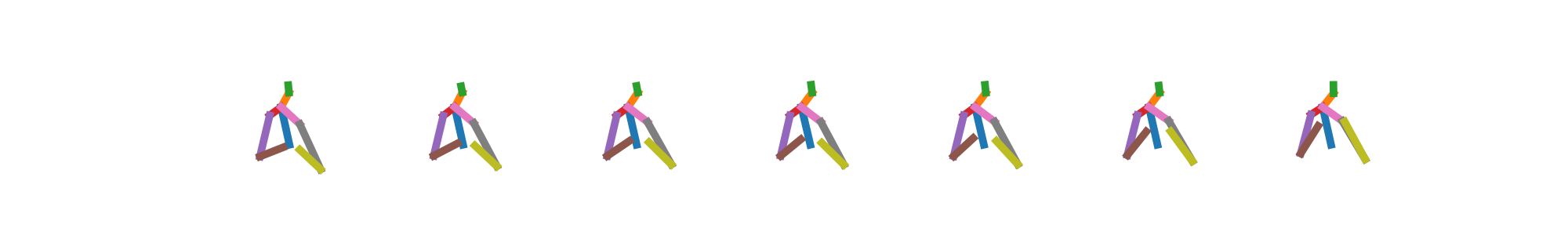}}\\
    \imagetop{(c)} & \imagetop{\includegraphics[trim=200 50 280 50 ,clip,width=0.87\linewidth]{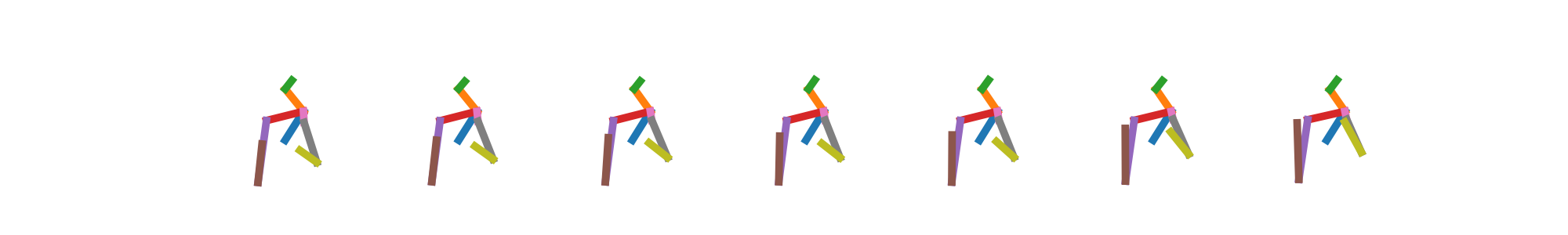}}\\
    \imagetop{(d)} & \imagetop{\includegraphics[trim=200 50 280 50 ,clip,width=0.87\linewidth]{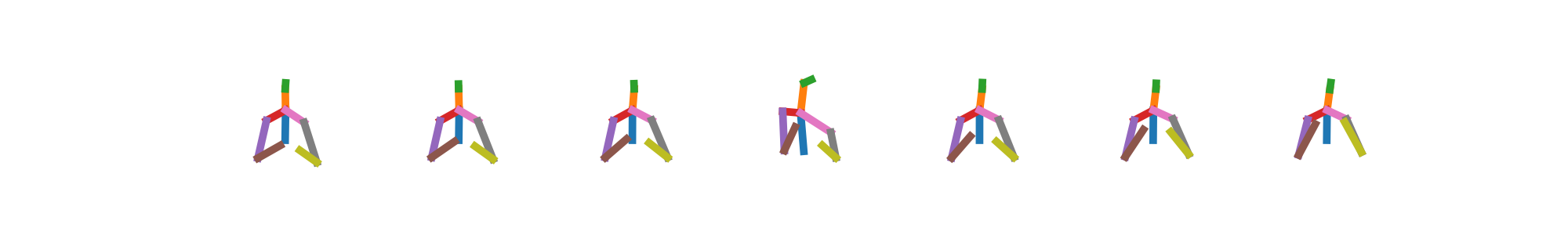}}\\
    \imagetop{(e)} & \imagetop{\includegraphics[trim=200 50 280 50 ,clip,width=0.87\linewidth]{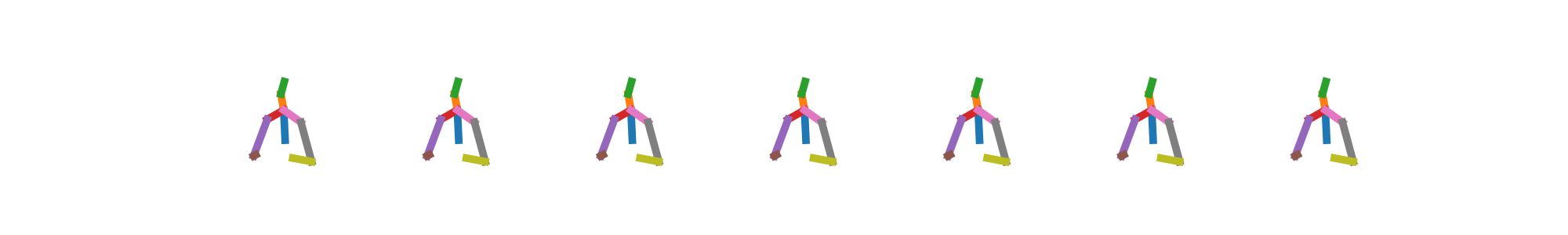}}\\
    \imagetop{(f)} & \imagetop{\includegraphics[trim=200 50 280 50 ,clip,width=0.87\linewidth]{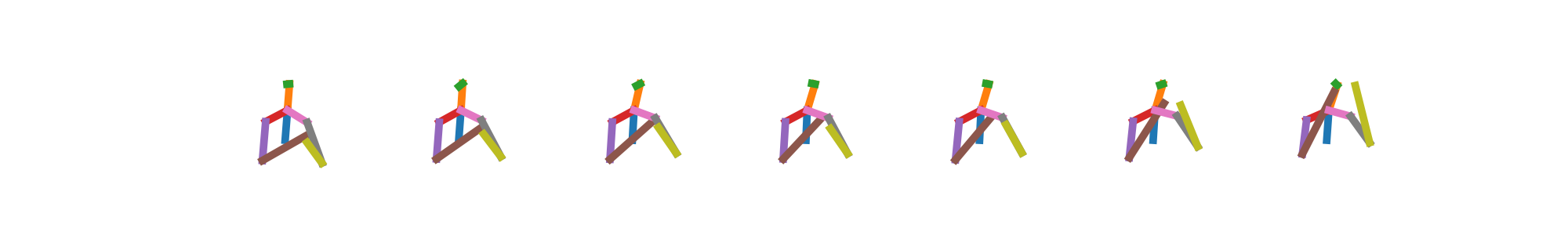}}\\
    \imagetop{(g)} & \imagetop{\includegraphics[trim=200 50 280 50 ,clip,width=0.87\linewidth]{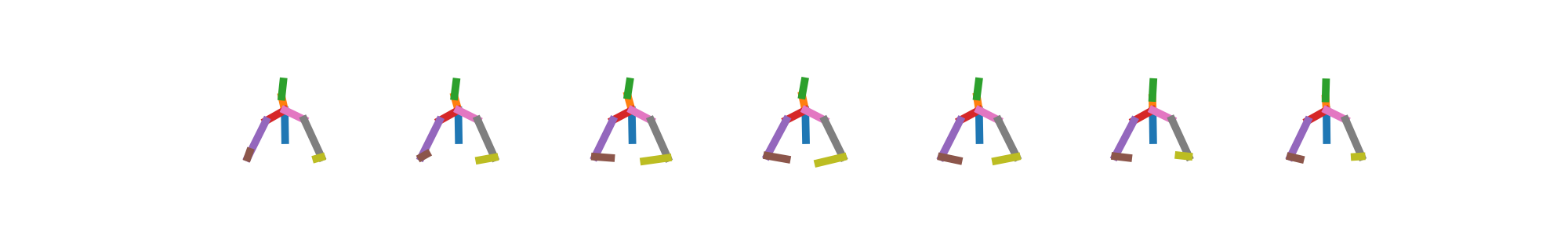}}\\
    \end{tabular}
    \caption{Samples of noisy gesture data to validate evaluation metrics. (a) None (original data), (b) Gaussian noise ($\zeta=0.001$), (c) Salt\&Pepper noise ($\zeta=0.1$), (d) Temporal noise ($\zeta=5$), (e--f) Multiplicative transformation in eigenposes ($\zeta=0.0, 2.0$), (g) Mismatched sample}
    \label{fig:noise}
\end{figure}

We explored the properties of the proposed FGD metric using synthetic noisy data. Five types of noisy data were considered. Gaussian noise and Salt\&Pepper (S\&P) noise were added to the joint coordinates of poses; the same noise data were added to all poses in a sequence, so that there is no artificial temporal discontinuity. Temporal noise was simulated by adding Gaussian noise to only a few time frames. Multiplicative transformation in ``eigenposes'' ${p}_i^{eigen}$ \cite{yoon2019robots} converted from ${p}_i$ using principal component analysis (PCA) was used to generate monotonous or exaggerated gestures. Mismatched gestures were also generated to examine how the metric responds to discrepancies between speech and gestures. The following shows how the noisy data were synthesized. The parameter $\zeta$ controls the overall disturbance levels. The dimension of a pose, $K$, is 30 (10 joints in 3D coordinates).

\begin{itemize}
  \item Gaussian noise: $\tilde{p}_i = p_i + x; x \sim \mathcal{N}_K(0,\,\zeta{}I)$
  \item Salt\&Pepper noise: $\tilde{p}_i = p_i + x$
      \begin{equation*}
        x_{k=1,...,K} =
        \begin{cases}
          0.2 & \text{if $u \leq \zeta/2; u \sim U(0,1)$}\\
          -0.2 & \text{if $\zeta/2 < u \leq \zeta$}\\
          0 & \text{otherwise}
        \end{cases}       
      \end{equation*}
  \item Temporal noise:
      \begin{equation*}
        \tilde{p}_i =
        \begin{cases}
          p_i + x; x \sim \mathcal{N}_K(0,\,0.003{}I) & \text{if } r \leq i < r+\zeta;\\ 
            & \text{$r$ is a random time step}\\
          p_i & \text{otherwise}
        \end{cases}       
      \end{equation*}
  \item Multiplicative transformation: $\tilde{p}_i^{eigen} = \zeta \cdot p_i^{eigen}$
  \item Mismatched samples: Select a fraction $\zeta$ of all samples and associate the input speech to random gestures in the TED test dataset to make mismatched samples.
\end{itemize}

\begin{figure*}
  \centering
  \includegraphics[width=\textwidth]{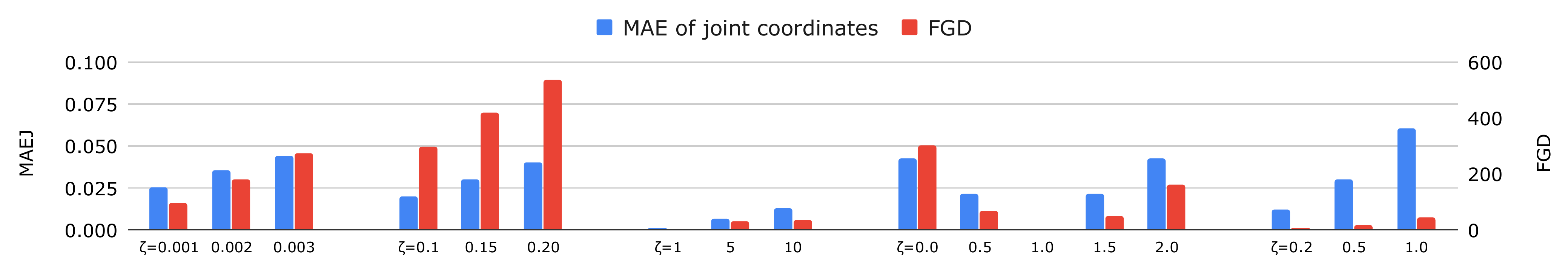}
  \begin{minipage}{\textwidth}
    \footnotesize
    \hspace{1.8cm}Gaussian noise\hspace{1.3cm}Salt\&Pepper noise\hspace{1.1cm}Temporal noise\hspace{1.3cm}Multiplicative transformation\hspace{1.5cm}Mismatched
  \end{minipage}
  \caption{Results of the metric validation experiment on the synthetic noisy dataset showing four types of noise. The disturbance level increases as $\zeta$ increases except for the multiplicative transformation. The disturbance level is lowest when $\zeta=1.0$ for the multiplicative transformation.}
  \label{fig:metric}
\end{figure*}

Figure \ref{fig:noise} shows samples of the synthetically noisy data. The Gaussian noise introduced changes across all joints, whereas the S\&P noise produces impulsive noise in a few joints. The temporal noise introduced discontinuities in motion. Multiplicative transformation was applied to eigenposes, so it controls the overall motion range. The mismatch noise shows a sample of nonmatching content and speech rhythms.

We measured FGD and mean absolute error of joint coordinates (MAEJ) which is calculated as MAE($\tilde{p}$, $p$). Figure \ref{fig:metric} shows the experimental results. For the Gaussian and S\&P noise, both FGD and MAEJ showed increasing distances as the disturbance level increases, but FGD showed larger distances for S\&P noise than Gaussian noise on average, unlike MAEJ. As shown in Figure \ref{fig:noise} (b) and (c), the sample with Gaussian noise still look like human poses, though with some distortions, whereas the sample with S\&P noise show unrealistic poses where the neck is out of the upper body. The samples with Gaussian noise are more perceptually plausible gestures than those with S\&P noise, so, in our view, having larger distances for S\&P noise is acceptable.

Both FGD and MAEJ showed relatively low values for the temporal noise even though discontinuous motion is perceptually unnatural. MAEJ calculates errors in each time frame independently, so it is obvious that MAEJ is not able to capture motion discontinuity. However, FGD, which encodes a whole sequence, also showed low distances unexpectedly. The primary reason is that the feature extractor used in FGD were not able to discriminate enough between the sequences with and without temporal noise. When we examined the reconstructed motion from the autoencoder, we found that the autoencoder tended to remove temporal noise.

For the multiplicative transformation, both metrics showed increasing distances as the disturbance level increased (larger or smaller than $\zeta=1.0$). MAEJ showed similar distances for $\zeta=0.0$ and 2.0, but FGD showed a much larger distance when $\zeta=0.0$. As shown in Figure \ref{fig:noise} (e) and (f), $\zeta=0.0$ and 2.0 make mean and exaggerated poses. If we consider several results and their diversity, exaggerated poses are perceptually favored over having the same mean poses regardless of input speech. Thus, it is reasonable to have larger distances for $\zeta=0.0$ for than 2.0, as FGD does.

Lastly, for the mismatched samples, both MAEJ and FGD showed increasing distances for more mismatched samples, but the increase in FGD was smaller than MAEJ. This result is not surprising since FGD considers a distribution formed by a set of gestures and is not aware of the input speech.

In this experiment, we found that FGD gives perceptually plausible results for the different gesture data of Gaussian noise, S\&P noise, and multiplicative transformation and has the limitation that it is not able to measure well enough temporal noise and match of speech and gestures. We found the characteristics of FGD; however, it is difficult to argue that the metric is suitable for use based on an experiment with synthetic data wherein only one human gesture example is used for each speech even though many-to-many mappings exist between speech and gestures. To further investigate the effectiveness of FGD and MAEJ, we compare these metrics to human judgements in the following section.

\section{User Study to Validate Evaluation Metrics} \label{sec:human_evaluation}

In this section, we validated FGD by comparing with subjective ratings from humans. We followed the overall experimental setting in the paper introducing Fr\'{e}chet video distance \cite{unterthiner2018towards}, but we had two separate user study sessions with 14 noise models and 10 trained gesture generation models. In the first session, 14 noise models (excluding Mismatched with $\zeta=0.2$ and 0.5) were used. In the second session, 10 gesture generation models showing different FGD were selected among models trained in the course of this study. We tried to select models equidistant in terms of FGD. The selected models are in different architectures, configurations, and training stages; see appendix B for the complete list of the selected models and their configurations. We also included the human gesture in both sessions.

We made videos showing a dummy character making gestures for each model. In the evaluation, pairwise preference comparisons were used instead of a Likert-scale rating. Co-speech gestures are subtle, so participants would have struggled to rate them on a five- or seven-point scale. Using pairwise preference comparisons reduces participants' cognitive load and yields reliable results, as discussed in \cite{clark2018rate}. The participants watched two videos of two different models and responded to one of three questions asking about their preference, human-likeness of motion, and speech--gesture matching: 1) ``Which gesture motion do you prefer?,'' 2) ``Which gesture motion is more natural and human-like?,'' and 3) ``Which gesture motion is more appropriate with the speech audio and words?'' The answer options were ``Video A,'' ``Video B,'' and ``Undecidable.'' For the question on human-likeness of motion, the videos were played without speech audio to make the participants assess only the motion. Each participant was asked to answer one randomly selected question in all of his/her trials, since the three questions are substantially correlated and the participants are prone to give the same answer if we ask three questions at the same time.

For the evaluation, speech samples with lengths of 5--10 s were drawn randomly from the TED test dataset. We only reviewed the quality of the extracted 3D human poses of the samples to exclude faulty samples that may mislead the performance of human gestures (the top line). Thirty speech samples were used in the evaluation after excluding four faulty samples where the speaker is manipulating an object, sitting on a chair, and occluded by a podium. Two models were randomly selected for each pairwise comparison to eliminate ordering effects.

Native or bilingual English speakers were recruited from Amazon MTurk. Each participant responded to 30 pairwise comparisons which were chosen randomly among all possible pairwise combinations (30 sentences $\times$ \{${}_{15}C_2$ or ${}_{11}C_2$\} $\times$ 3 questions). They took 15--30 min to complete the task, and 2.5 USD was given as a reward. We also included an attention check presenting two copies of the same video side by side. The participants who did not answer ``undecidable'' in this case were excluded. In the first session with the noise models, a total of 28 subjects participated, but we analyzed the results from 22 subjects after excluding six subjects failed the attention check. There were 13 male and 9 female subjects, and they were 36.9 ± 11.5 years old. In the second session with the trained generation models, a total of 51 subjects participated and 21 subjects were excluded. There were 15 male and 15 female subjects, and they were 42.8 ± 13.2 years old. The total numbers of answers were 660 and 900 at the first and second session, respectively.

\definecolor{Gray}{gray}{0.9}
\newcolumntype{P}[1]{>{\centering\arraybackslash}p{#1}}

\begin{table}
\caption{Agreement of the evaluation metric to human judgements in the user study on (a) noise models and (b) trained gesture generation models. We also report the agreements between human subjects as a top line. Higher numbers are better.}
\label{tab:agreement}
\centering
\begin{tabular}{p{0.30\linewidth}P{0.13\linewidth}P{0.15\linewidth}P{0.24\linewidth}} 
\toprule
& \multicolumn{3}{c}{Agreement (\%)}                        \\ 
\cmidrule{2-4}
Metric & Preference & Human-likeness of motion & Speech--gesture match  \\ 
\midrule
\rowcolor{Gray}
\multicolumn{4}{l}{(a) Noise models}\\ 
\midrule
MAE of joint coordinates (MAEJ) & 50.9 & 55.9 & 60.5 \\
MAE of acceleration & 46.5 & 47.7 & 46.3 \\
FGD & 64.8 & 63.6 & 66.3 \\
\midrule
Between human subjects & 83.3 & 72.2 & 85.7 \\ 
\midrule
\rowcolor{Gray}
\multicolumn{4}{l}{(b) Trained gesture generation models}\\ 
\midrule
MAE of joint coordinates (MAEJ) & 37.7 & 48.2 & 32.8 \\
MAE of acceleration & 34.9 & 40.0 & 38.9 \\
FGD & 70.5 & 59.6 & 70.2 \\
\midrule
Between human subjects & 73.1 & 78.8 & 94.4 \\ 
\bottomrule
\end{tabular}
\end{table}

We evaluated the objective evaluation metrics by comparing those with human judgements, and the results are shown in Table \ref{tab:agreement}. MAE of acceleration was used to assess dance motion \cite{aristidou2015folk} and gestures \cite{kucherenko2019analyzing}, and it focuses on motion rather poses. The agreement values were calculated as the number of comparisons in which each metric agreed human judgement divided by the total number of comparisons. ``Undecidable'' responses were not included in the analysis. In both sessions, FGD showed greater agreement with human judgements than did MAE of joint coordinates and MAE of acceleration on all questions. However, FGD was performed less than the agreements between humans; in particular, FGD showed the lowest agreement of 53.5\% for temporal noise as discussed in Section 5.2. 

\begin{figure}
  \centering
  \includegraphics[width=\linewidth]{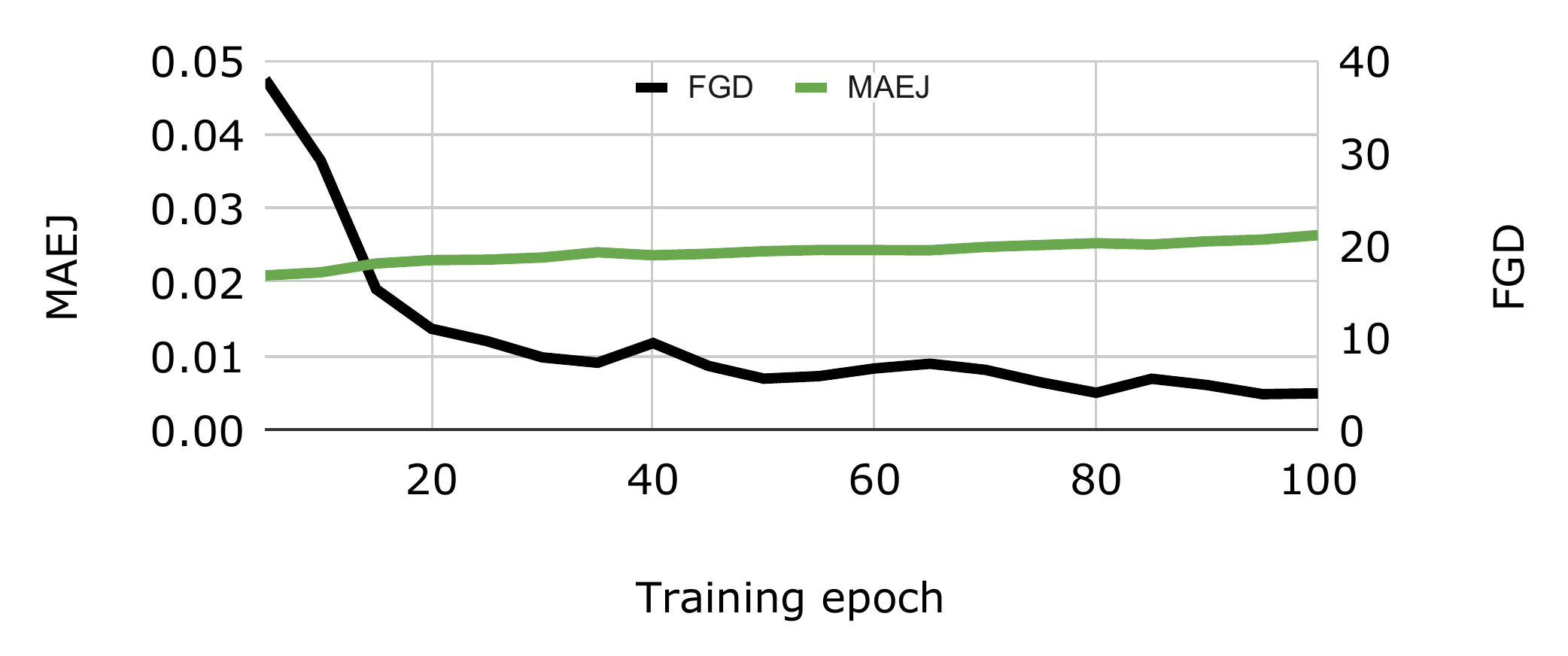}
  \caption{Validation learning curves measured by mean absolute error of joint coordinates (MAEJ) and Fr\'{e}chet gesture distance (FGD).}
  \label{fig:train_curve}
\end{figure}

By considering both experimental results on synthetic noisy data and human judgements, FGD is a plausible objective metric. In addition, when we examine learning curves, which are shown in Figure \ref{fig:train_curve}, FGD showed a decreasing trend when the distribution of generated gestures becomes more similar to the reference distribution as training continues. In contrast, MAEJ showed a flat learning curve. The lowest MAEJ is at Epoch 6, in which only static mean poses appear for all speech contexts. In the following experiments, we use FGD to compare models.

All subjects were asked to write the reasons for their selection. Most of them said they preferred gestures that were fit to speech words and audio, as we had assumed in the present paper. Opinions on gesture dynamics were mixed. Some participants liked dynamic or even exaggerated gestures, whereas some other participants preferred moderate gestures with a few large movements for emphasis. This implies that the gesture styles must be adapted as per the users' preference.
\section{Experiments and Human Evaluation} \label{sec:experiment}

\subsection{Qualitative Results}

\begin{figure*}
  \centering
  \includegraphics[width=\textwidth]{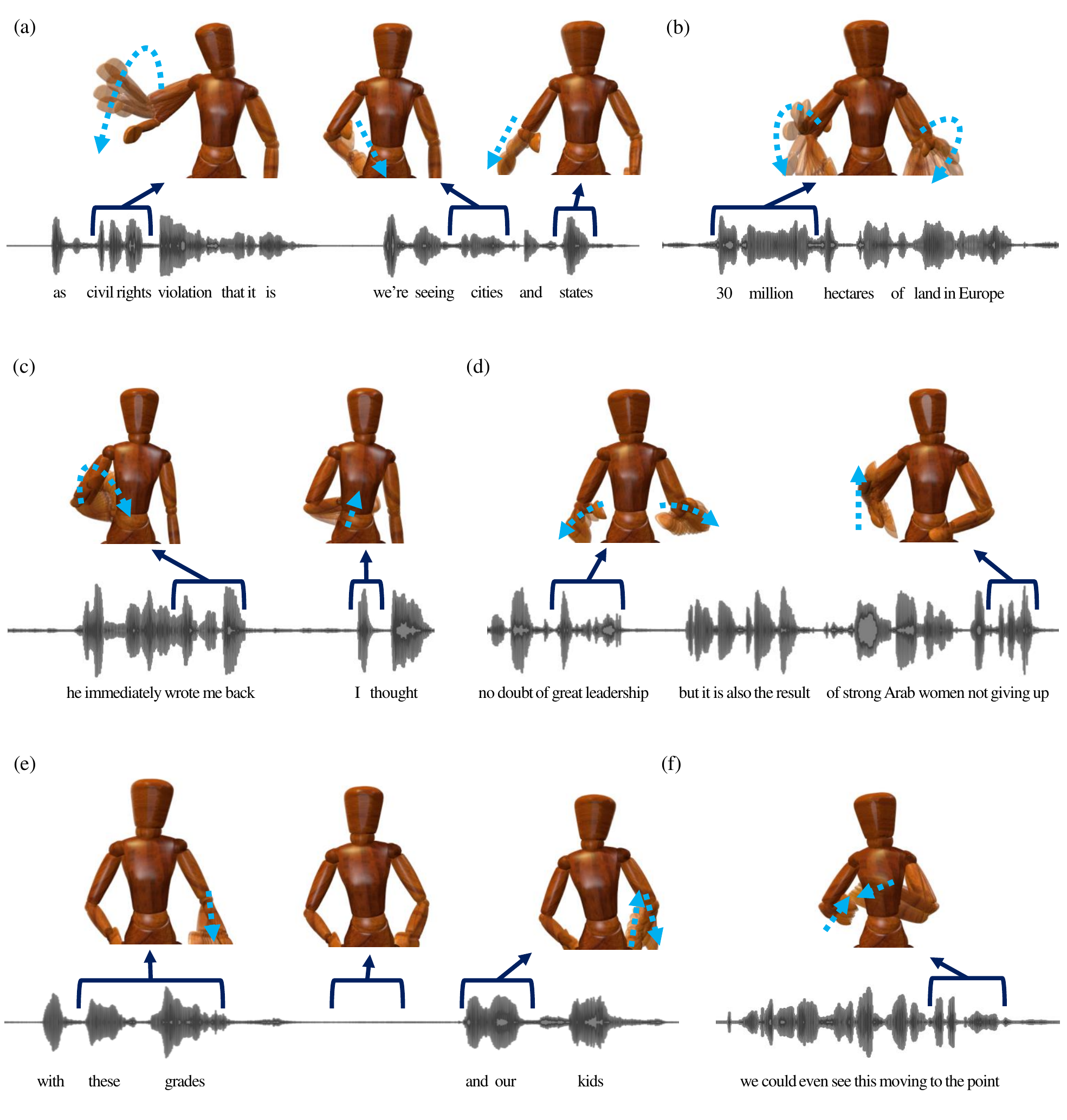}
  \caption{Sample results of co-speech gesture generation from the trimodal speech context of text, audio, and speaker identity. Motion history images for some parts are depicted along with the speech text and audio signals. In (a), the character makes metaphoric gestures when saying ``civil rights'' and beat gestures for ``cities and states.'' In (b) and (d), there are metaphoric gestures for the words of ``30 million,'' ``great leadership,'' and ``giving up.'' In (c), a deictic gesture appears when the character says ``I.'' In (e), we can find the character does not gesture in the middle of the silence. An iconic gesture is also found in (f).}
  \label{fig:samples}
\end{figure*}

Figure \ref{fig:samples} shows the gesture generation results for the speech in the test set of the TED gesture dataset. The gestures are depicted using a 3D dummy character. The poses represented as directional vectors were retargetted to the character with fixed bone lengths, and the gesture sequences were upsampled using cubic spline interpolation to 30 FPS. We used the same retargeting procedure for all animations. The character makes metaphoric gestures when saying ``civil rights,'' ``30 million,'' or ``great leadership.'' An iconic gesture also found for the words ``to the point.'' Gesture generation depends on speech rhythm and presence or absence of speech as shown in the sample (a) and (e). A deictic gesture also appears in (c) when the character says ``I.'' Please see the supplementary video for the animated results.

\subsection{Comparisons with state-of-the-art models} \label{sec:sota}

We compared the proposed model with three models from previous studies. The first model compared is attentional Seq2Seq, which generates gestures from speech text \cite{yoon2019robots}. We followed the original implementation provided by the authors but the gesture representation was modified to be identical to the proposed model. The second comparison model is Speech2Gesture \cite{ginosar2019gestures}, which generates gestures from speech audio using an encoder--decoder neural architecture and learns to generate human-like gestures by using an adversarial loss during training. Spectrograms were used to represent audio in this model. The third one is the joint embedding model \cite{ahuja2019language2pose}, which creates human motion from motion description text. This model maps text and motion to the same embedding space. We embedded the input speech text and audio together to the same space as the motion. The same encoders in our model were used to process the audio and text, and 4-layered GRUs were used for gesture generation. All models were trained on the same TED dataset for the same number of epochs. We modified the original architectures of the baselines to generate the same number of poses (i.e., 30) and to use four seed poses for consecutive syntheses. The learning rate and weights of loss terms in the baselines were optimized via grid search for best FGD. 

\begin{figure}
  \centering
  \begin{tabular}{@{}l@{}l@{}} 
    \multicolumn{2}{c}{\includegraphics[width=0.95\linewidth]{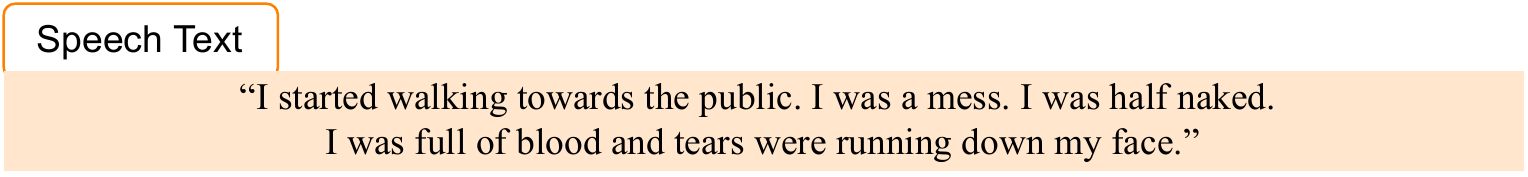}}\\
    \multicolumn{2}{c}{\includegraphics[width=0.95\linewidth]{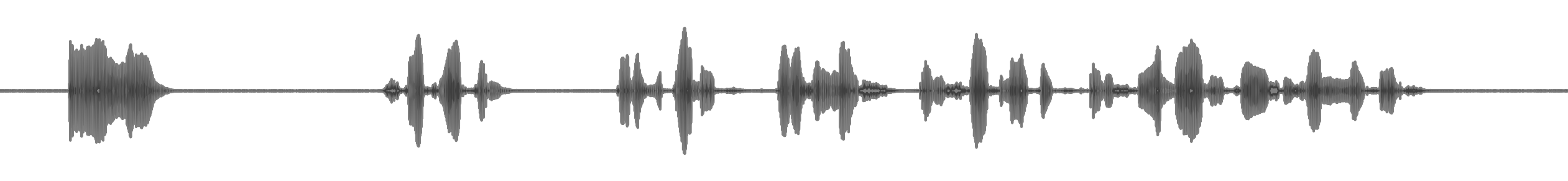}}\\
    \imagetop{(a)}&\imagetop{\includegraphics[width=0.9\linewidth]{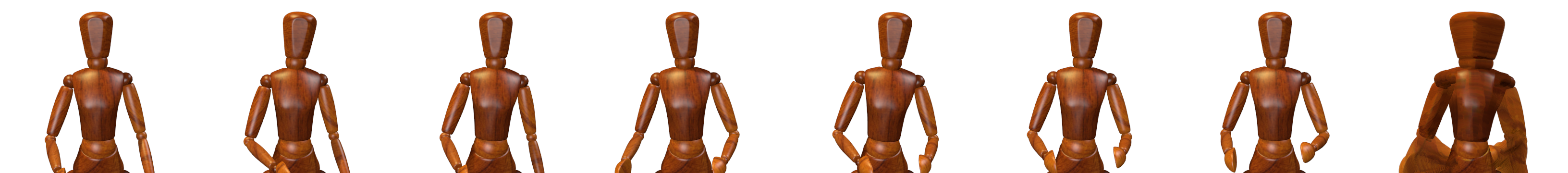}}\\
    \imagetop{(b)}&\imagetop{\includegraphics[width=0.9\linewidth]{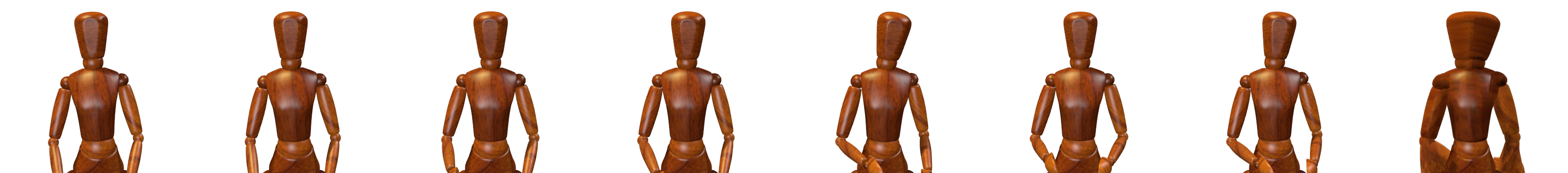}}\\
    \imagetop{(c)}&\imagetop{\includegraphics[width=0.9\linewidth]{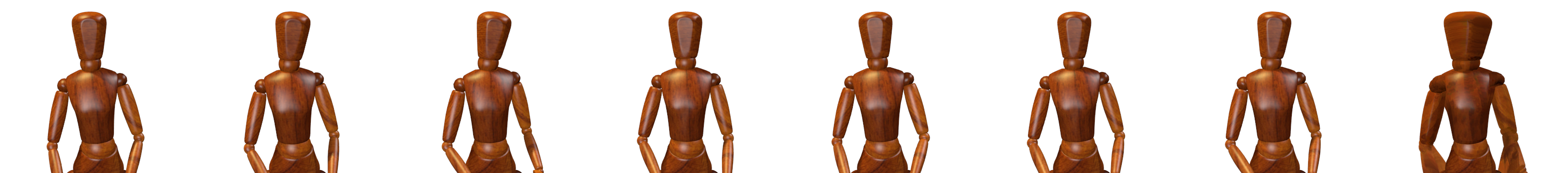}}\\
    \imagetop{(d)}&\imagetop{\includegraphics[width=0.9\linewidth]{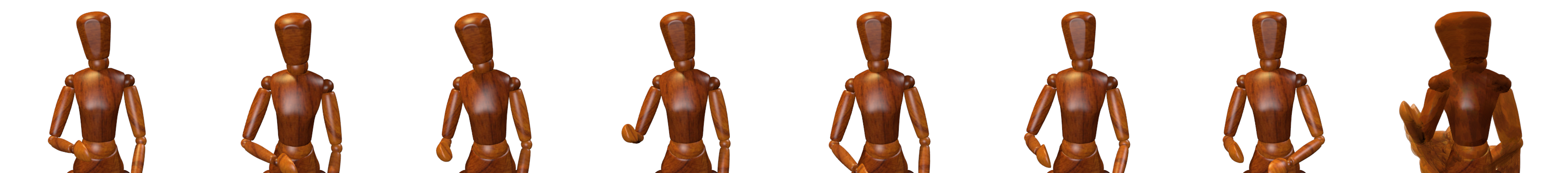}}\\
  \end{tabular}
  \caption{Sample results of (a) attentional Seq2Seq, (b) Speech2Gesture, (c) joint embedding, and (d) the proposed model for the same input speech. Seven evenly sampled frames are shown for the resulting pose sequences. The last column shows motion history images in which all frames are superimposed. Please see the supplementary video for animated results.}
  \label{fig:sota}
\end{figure}

Figure \ref{fig:sota} shows sample results from each model for the same speech. The joint embedding model generated very static poses, failing to learn gesticulation skills. The relationship between speech and gestures are weak and subtle, making it difficult to map speech and gestures to a joint embedding space. All other models generated plausible motions, but there were differences depending on the modality and training loss considered. Attentional Seq2Seq generated different gestures for different input speech sentences, but the motion tended to be slow and we found a few discontinuities between the seed poses and generates poses. The Speech2Gesture model used an RNN decoder similar to attentional Seq2Seq, but it showed better motion with the help of its adversarial loss component. However, because it uses only a single speech modality, audio, Speech2Gesture generated monotonous beat gestures. The proposed model successfully generated large and dynamic gestures as shown in the supplementary video.

\begin{table}
\caption{Results of comparisons with state-of-the-art models. Lower numbers indicate better performance (Bold: \textbf{best}, Underline: \underline{second}).}
\label{tab:compare}
\centering
\begin{tabularx}{\linewidth}{Xc}
  \toprule
  Method & FGD\\ 
  \midrule
  Attentional Seq2Seq \cite{yoon2019robots} & \underline{18.154}\\
  Speech2Gesture \cite{ginosar2019gestures} & 19.254\\
  Joint embedding model \cite{ahuja2019language2pose} & 22.083\\
  Proposed & \textbf{3.729}\\
  \bottomrule
\end{tabularx}
\end{table}

\begin{figure}
  \centering
  \includegraphics[width=\linewidth, trim = 2cm 5cm 2cm 4cm, clip]{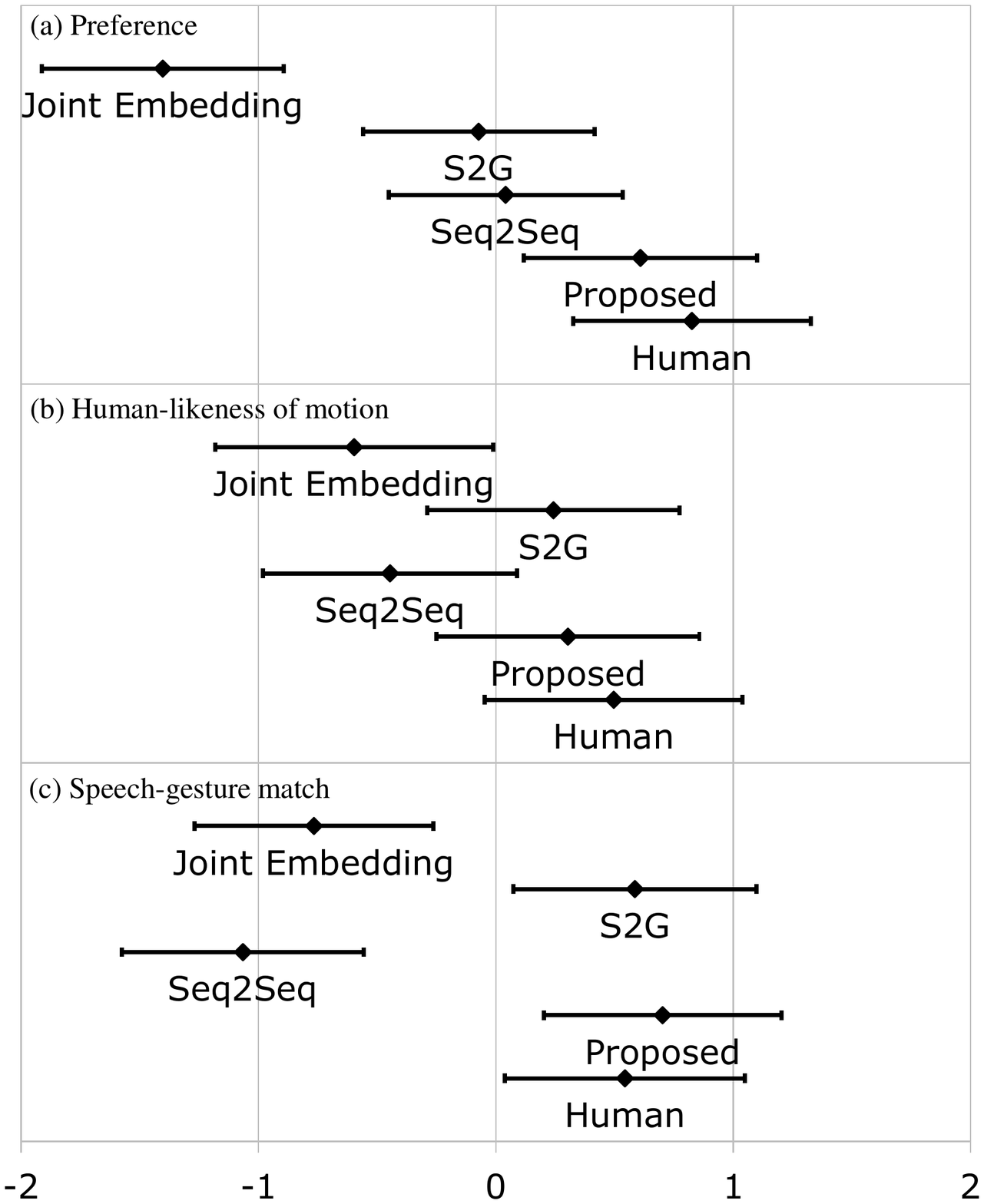}  
  \caption{The results of human evaluation for the three questions about (a) preference, (b) human-likeness of motion, and (c) speech--gesture match. The ranking is calculated using the Bradley-Terry model and the horizontal axis represents the winning probability against the other methods. Mean and standard deviation are depicted through Bayesian inference for the Bradley-Terry model \cite{chu2005extensions}. S2G denotes the Speech2Gesture method.}
  \label{fig:ut_results}
\end{figure}

The proposed model performed the best in terms of FGD (Table \ref{tab:compare}). We also analysed the human evaluation results by computing ranks from pairwise comparisons using the Bradley---Terry model \cite{chu2005extensions}. Pairwise comparisons were collected from another 14 MTurk subjects that passed the same attention check as before. The same settings described in Section \ref{sec:human_evaluation} were used, but only the four models in Table \ref{tab:compare} and human gestures were compared. Figure \ref{fig:ut_results} shows the results. For all the questions, the proposed method achieved better results than Attentional Seq2Seq, Speech2Gesture, and joint embedding methods, but the differences between the proposed method and Speech2Gesture were not distinct in the the human-likeness of motion and speech--gesture match questions. We also tested statistical significance between the proposed method and the others by using the Chi-Square Goodness of Fit test over the null hypothesis that the probabilities of the pairwise choices are equal to 50\% (the choice of ``undecidable'' was not counted). In the preference, the difference between the proposed and joint embedding method was significant (p < 0.01). In the speech--gesture match, Seq2Seq and joint embedding methods were significantly different from the proposed method (p < 0.01 and < 0.05, respectively).

The proposed method showed better results than the previous methods objectively and subjectively. Also, the proposed method is mostly tied with human gestures in the user study. This indicates the superiority of the proposed method, but we cannot conclude that the proposed method performed equally well as humans since the human gestures used in the experiments were based on automatically extracted poses from TED videos and all motion was retargetted to a restricted character without face or hand expressions.

\subsection{Ablation Study}

\begin{table}
\caption{Results of the ablation study for the proposed model. Lower numbers are better. Ablations are not accumulated.}
\label{tab:ablation}
\centering
\begin{tabularx}{\linewidth}{Xc}
  \toprule
  Configuration & FGD\\ 
  \midrule
Proposed (no ablation) & 3.729\\
Without speech text modality & 4.701\\
Without speech audio modality & 4.874\\
Without speaker ID & 6.275\\
Without adversarial scheme & 9.712\\
Without regularization terms $L_G^\textrm{style}$ and $L_G^\textrm{KLD}$ & 5.756\\
  \bottomrule
\end{tabularx}
\end{table}

An ablation study was conducted to understand the proposed model in detail. We eliminated components from the proposed model that was used in the comparison with the state-of-the-art models. Table \ref{tab:ablation} summarizes the results of the ablation study. Removing each modality of text, audio, and speaker ID reduced the model's performance; this shows that all three modalities used in the proposed model had positive effects on gesture generation. Among the loss terms, removing the adversarial term and regularization terms also worsened FGD. In particular, when we trained the model without the adversarial scheme, the model tended to generate static poses close to the mean pose.

Although, when ablating different modalities, excluding the speaker ID degraded the FGD the most, we could not find a noticeable degradation in our subjective impression of motion quality than ablating text or audio modalities. In our view, this is attributed to that overall diversity was reduced without the divergence regularization $L_G^\textrm{style}$ and that the property of FGD that measures not only motion quality but also diversity. There is no concrete way to disentangle the factors of quality and diversity in FGD as well as FID. However, we hypothesise that the covariance matrix of the fitted Gaussian is more related to the diversity than to quality. The trace of the covariance matrix was 244, which is less than that of the human gestures and of the models without the text or audio modalities (299, 258, and 250, respectively). This indirectly suggests that generated gestures were less diverse without speaker IDs and $L_G^\textrm{style}$.

\begin{figure}
  \centering
  \includegraphics[width=\linewidth]{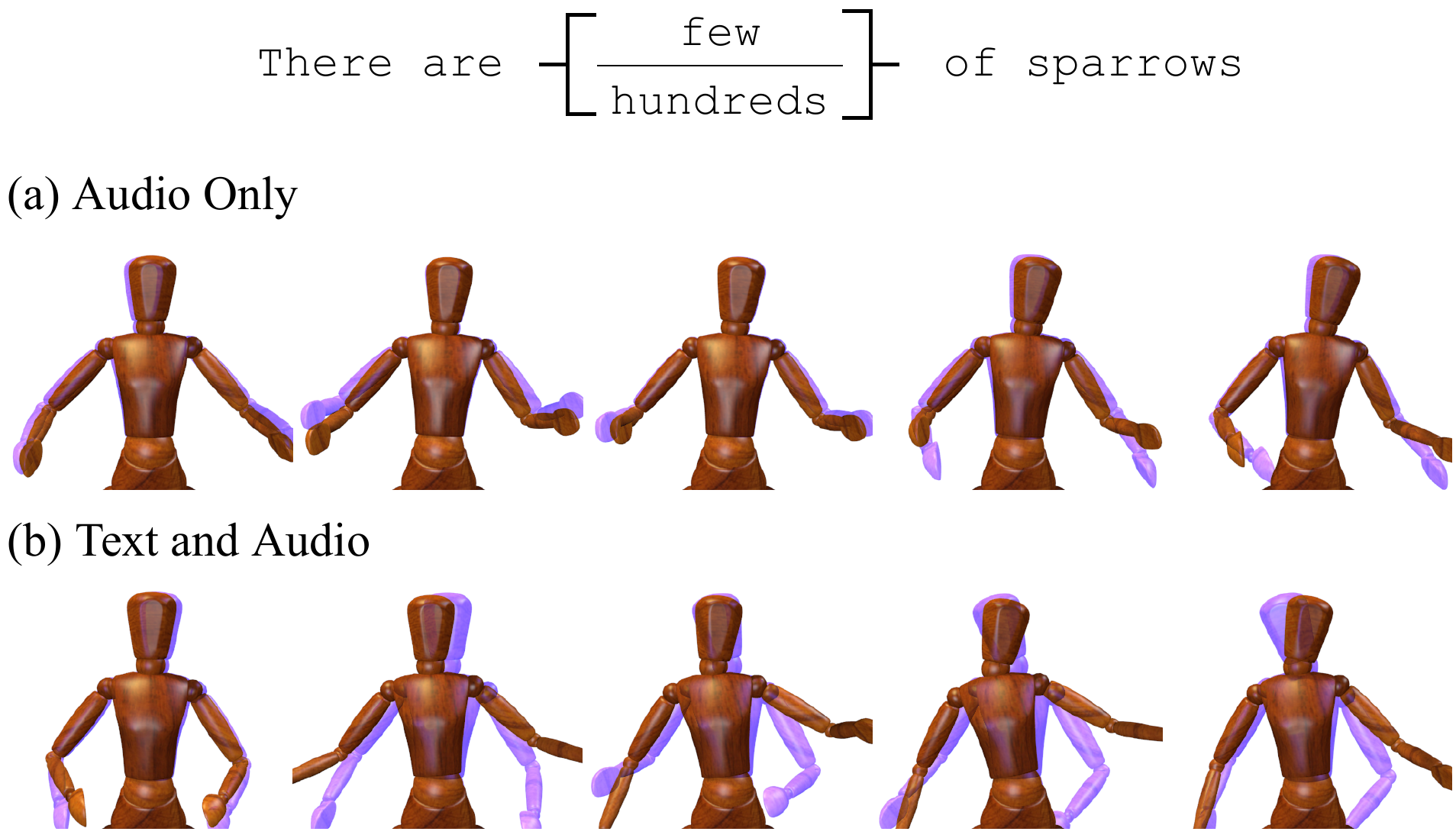}
  \caption{Visualization of how a gesture changes when a word is changed in a sentence. We compare the results of (a) the ablated model without text modality and (b) proposed model considering both text and audio. The generated gesture for the original and altered sentences are overlaid for five evenly sampled frames. When we consider both text and audio, the model generates more different gestures for the changes in speech content.}
  \label{fig:text_altering}
\end{figure}

The text modality had the least effect on FGD. In the proposed model, speech text and audio are treated as independent modalities; however, strictly speaking, audio contains text information because we can transcribe text from audio. Although the above ablation study showed that the FGD worsened without the text modality, it was less significant than excluding audio or speaker IDs. We further verified the effect of the text with an additional experiment. Figure \ref{fig:text_altering} shows how the generated gestures differed when a word was altered in the input speech with the model considering both text and audio and the model considering only audio. Although the model considering both text and audio generated different gestures (widening arms) when the word ``hundreds'' replaced the words ``few,'' there was only a slight change in motion when we used the audio-only model. We synthesized speech audio using Google Cloud TTS \cite{googletts} for both original and altered text.

We also conducted the above text-altering experiment quantitatively. For 1,000 samples randomly selected from the validation set, a word in a speech sentence was changed to a synonym or antonym taken from WordNet \cite{miller1995wordnet}. If there were several synonyms or antonyms, the one closest in duration to the original word was selected to minimize the change in the length of the speech audio. Synthesized audio was used and the experiment was repeated 10 times due to the randomness in selecting samples and words. We report the FGD between the generated samples before and after text alteration; this measure is unlike all other FGD measures, which compare human motion and generated motion, in the paper. The model considering text and audio (2.433 ± 0.483) showed a significantly higher FGD than the model considering only audio (1.604 ± 0.275) (paired t-test, p < 0.001), indicating that using text and audio modalities together helps to generate diverse gestures according to the changes in the speech text. This argument is also backed by the result that the FGD when a word was replaced by an antonym (2.567 ± 0.484) was significantly higher than when replaced by a synonym (2.299 ± 0.467) (paired t-test, p < 0.05) in the model using both text and audio.

\subsection{Incorporating Synthesized Audio}

\begin{figure}
  \centering
  \begin{tabular}{@{}l@{}l@{}l@{}} 
    \multicolumn{3}{c}{\includegraphics[width=0.95\linewidth]{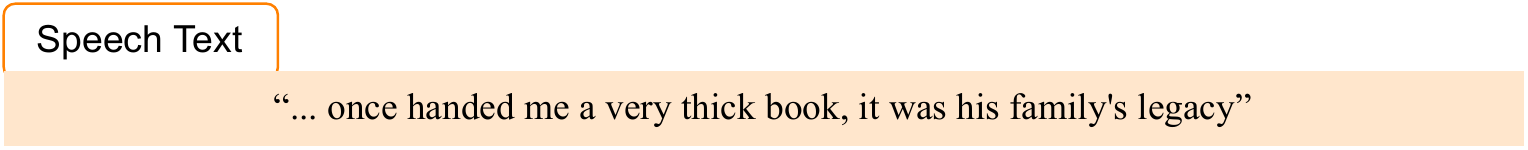}}\\
    
    \multirow{2}{1.1em}{\imagetop{(a)}}&\imagetop{\includegraphics[width=0.72\linewidth]{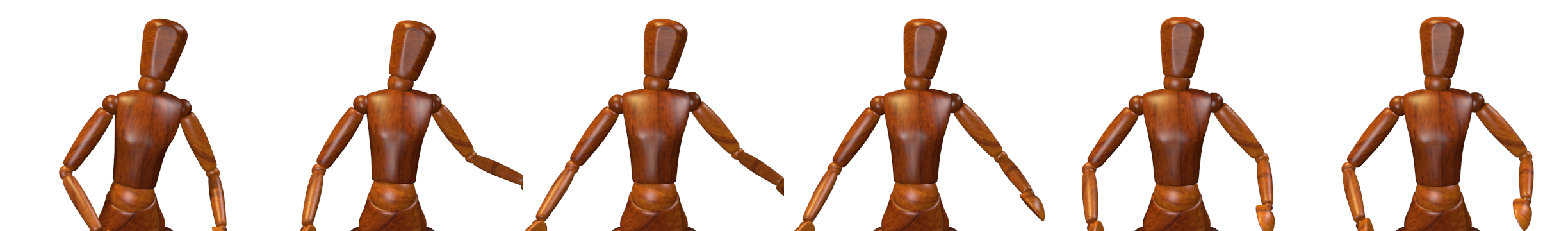}}&\multirow[c]{2}{*}[0.7em]{\imagetop{\includegraphics[trim=0 0 0 20,clip,width=6em]{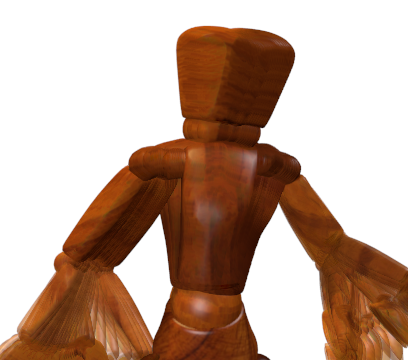}}}\\
    &\imagetop{\includegraphics[width=0.72\linewidth]{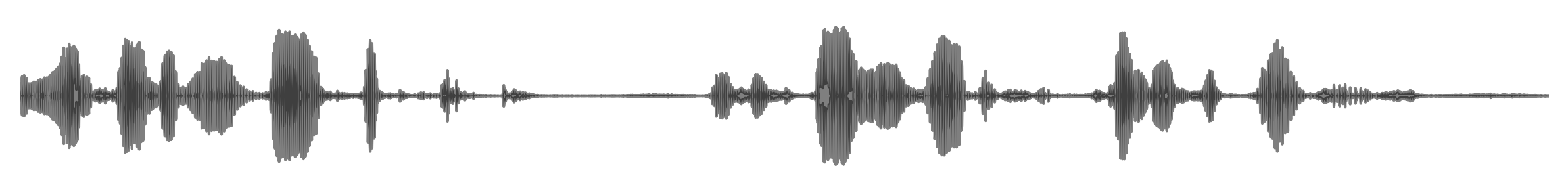}}&\\
    
    \multirow{2}{1.1em}{\imagetop{(b)}}&\imagetop{\includegraphics[width=0.72\linewidth]{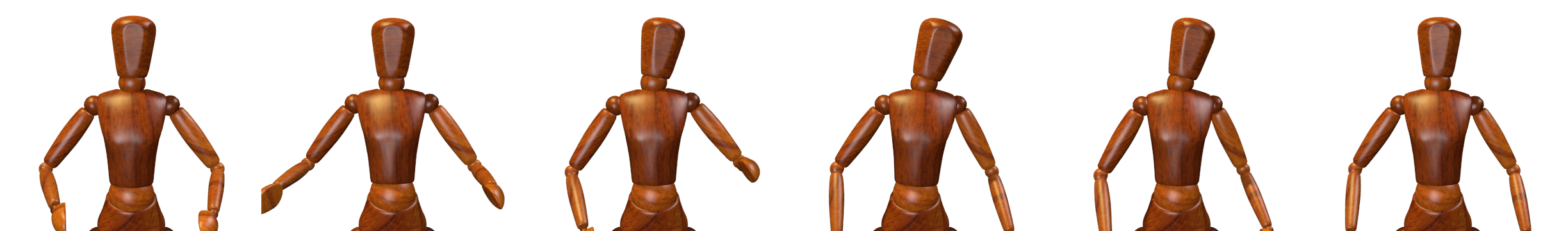}}&\multirow[c]{2}{*}[0.7em]{\imagetop{\includegraphics[trim=0 0 0 20,clip,width=6em]{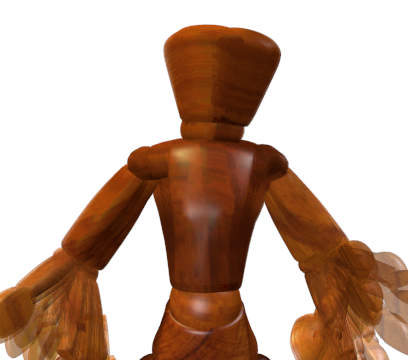}}}\\
    &\imagetop{\includegraphics[width=0.72\linewidth]{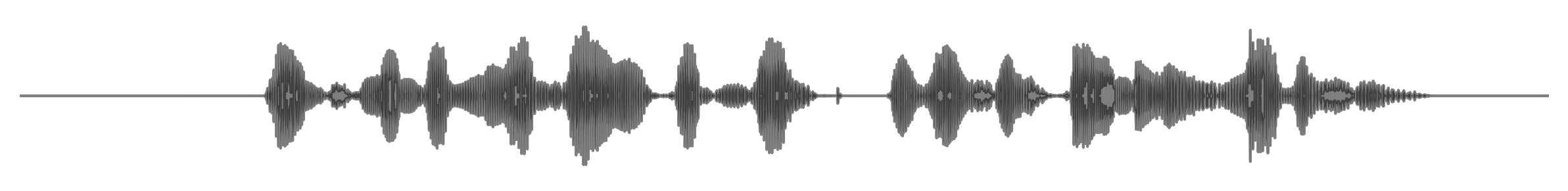}}&\\
    
    \multirow{2}{1.1em}{\imagetop{(c)}}&\imagetop{\includegraphics[width=0.72\linewidth]{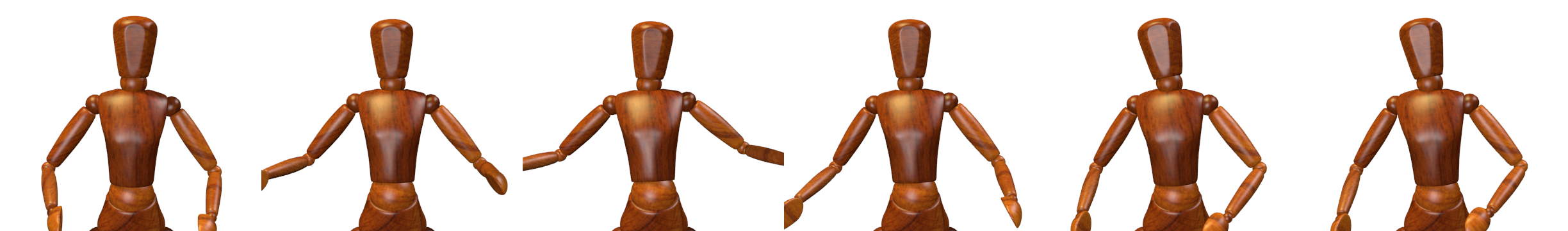}}&\multirow[c]{2}{*}[0.7em]{\imagetop{\includegraphics[trim=0 0 0 20,clip,width=6em]{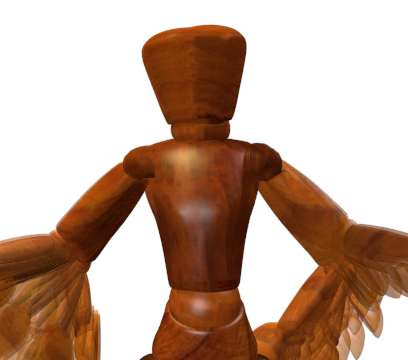}}}\\
    &\imagetop{\includegraphics[width=0.72\linewidth]{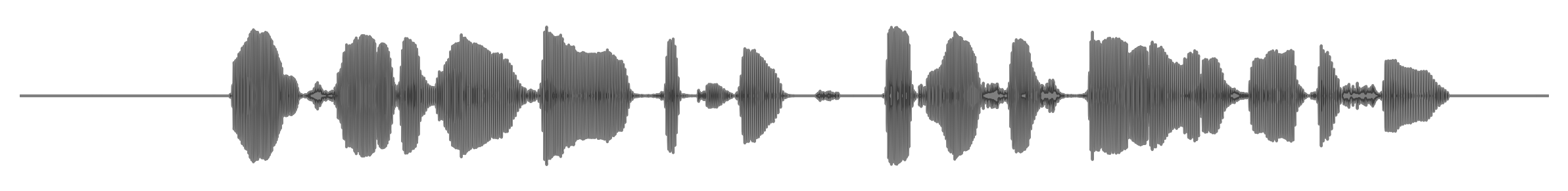}}&\\
    
    \multirow{2}{1.1em}{\imagetop{(d)}}&\imagetop{\includegraphics[width=0.72\linewidth]{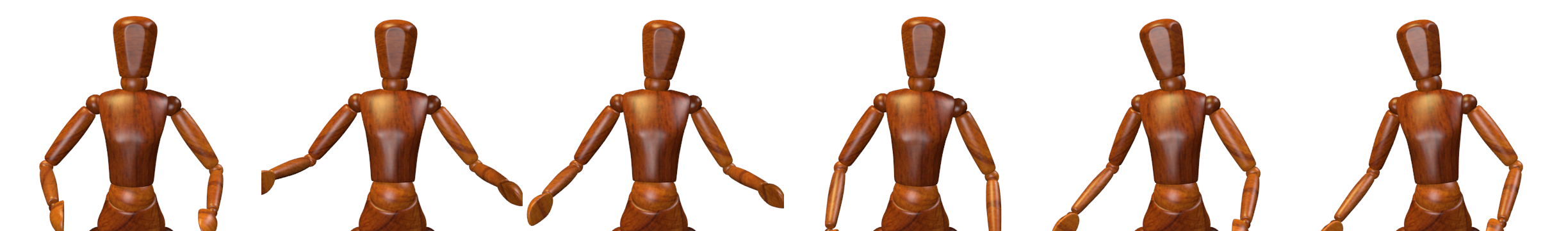}}&\multirow[c]{2}{*}[0.7em]{\imagetop{\includegraphics[trim=0 0 0 20,clip,width=6em]{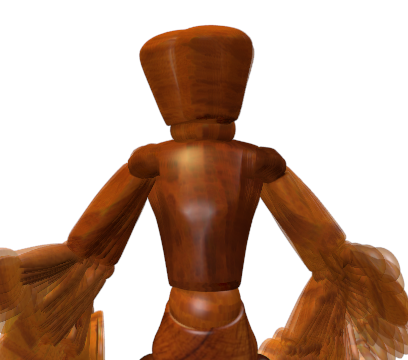}}}\\
    &\imagetop{\includegraphics[width=0.72\linewidth]{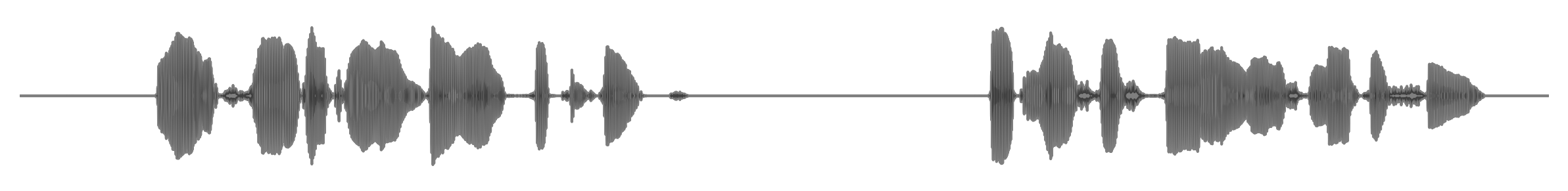}}&\\
  \end{tabular}
  \caption{Co-speech gesture generation results with (a) original human speech audio, (b) synthesized audio of a male voice, (c) synthesized audio of female voice, and (d) synthesized audio of a female voice with pauses. The proposed model can generate gestures from synthesized audio of different voices and rhythm.}
  \label{fig:tts}
\end{figure}

Many artificial agents use synthesized audio since recording a human speaking for every word is infeasible. We tested that the proposed model, trained with human speech audio, also can work with synthesized audio. Figure \ref{fig:tts} and the supplementary video shows some results using synthesized audio with different voices. Google Cloud TTS \cite{googletts} was used in this experiment. The proposed model worked well with synthesized audio of different voices, prosody, speed, and pauses. When the speech is fast, the model generates rapid motion. The model also reacts to inserted speech pauses by generating static poses for the silence period. 
  
\subsection{Analysis of the Learned Style Embedding Space} \label{section:speaker_embed}

\begin{figure*}
  \centering
  \includegraphics[width=0.95\textwidth]{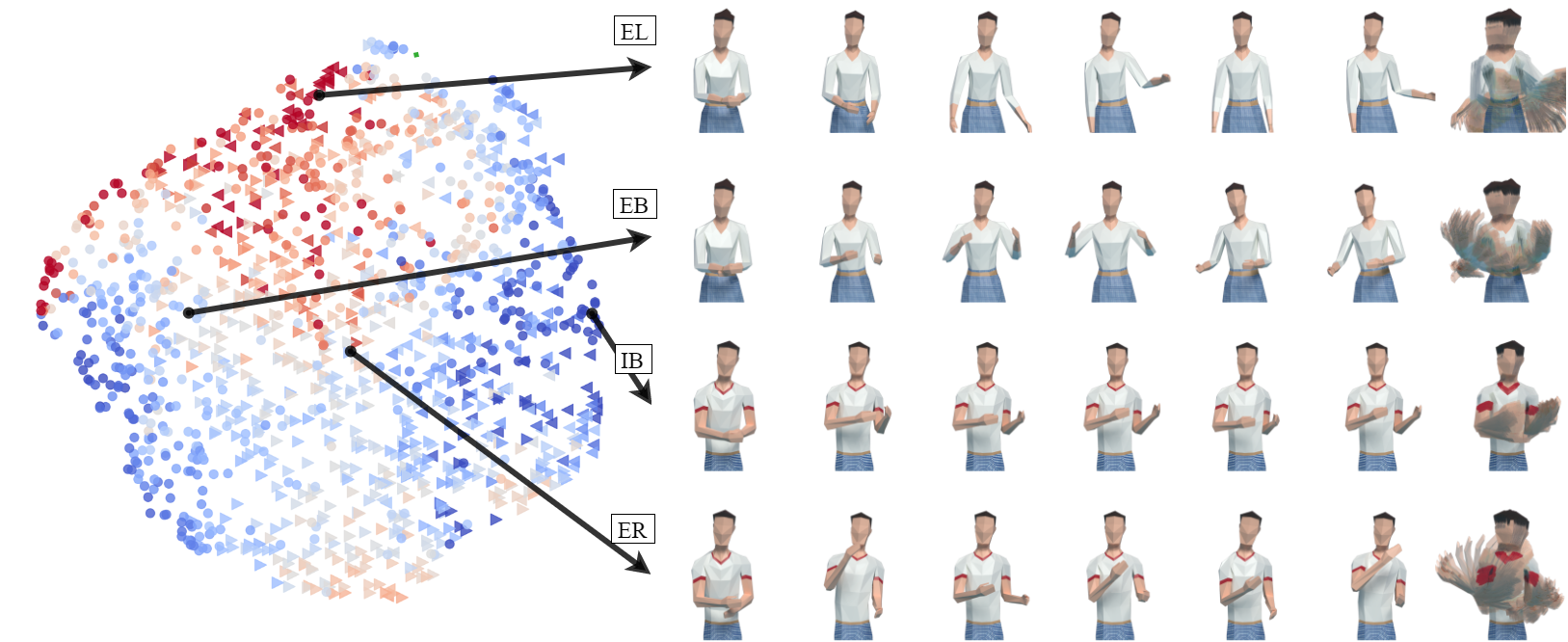}
  \caption{Visualization of the style embedding space and sample generation results for the different style vectors. All speaker identities are mapped to style feature vectors $f^\textrm{style}$, and we visualize the feature vectors in two dimensions by using UMAP \cite{mcinnes2018umap}. The points represent degrees of motion variance via color and degree of handedness by its marker types. We labeled the sampled style vectors according to the overall motion variance and handedness as `IB' for the introvert style of moving both hands similarly, `ER' for the extrovert style of moving the right hand more, and so on. All gesture results are generated from the same speech.}
  \label{fig:speaker}
\end{figure*}

The proposed model can generate different gesture styles for the same speech. Figure \ref{fig:speaker} visualizes the trained style embedding space and the gestures generated with different style vectors for the same input speech. To understand the style embedding space closely, we depict the motion statistics of the generated gestures for each style vector corresponding to speaker ID with the marker color and shape in the figure. Colors from red to blue correspond to higher and lower temporal motion variances. A larger motion variance can be called an extrovert style and the opposite is an introvert style. We also calculated the temporal motion variance for the right and left arms separately and used different marker shapes to indicate styles of handedness. Styles using the right and left arms more are depicted as $\blacktriangleright$ and $\blacktriangleleft$ respectively, and the rest are depicted as $\bullet$. As shown in Figure \ref{fig:speaker}, similar styles are clustered, and users can easily choose the desired style from the embedding space after traversing it.

\section{Conclusions and Limitations} \label{sec:discussion}

In this paper, we presented a co-speech gesture generation model that generates upper-body gestures from input speech. We proposed a temporally synchronized architecture using the three input modalities of speech text, audio, and speaker ID. The trained model successfully generated various gestures matching the speech text and audio; different styles of gestures could be generated by sampling style vectors from a style embedding space. A new metric, FGD, was introduced to evaluate the generation results. The proposed metric was validated using synthetic noisy data and measuring the agreement with human judgements. The proposed generation method showed better results than previous methods both objectively and subjectively as determined by the FGD metric and human evaluation. We also highlighted different properties of the proposed model through various experiments. The model can generate gestures with synthesized audio of various prosody settings. Additionally, the style embedding space was trained to be a continuous space where similar styles are distributed closely.

There is room for improvement in the present research. First, it is difficult to control the gesture generation process. Although style manipulation is possible, users are not able to set constraints on gestures. For example, we might want an avatar to make a deictic gesture when the avatar says a specific word. Most end-to-end neural models have this controllability issue \cite{jahanian2019steerability}. It would be interesting to extend the current model to have further controllability, for example, by adding constraining poses in the middle of generation. Second, FGD need to be improved. In non-verbal behavior, subtle motion is as important as large motion, but the feature extractor trained by motion reconstruction might fail to capture subtle motion. It is also necessary to separately evaluate motion quality and diversity for in-depth comparisons between generation models. Third, we only considered the motion of upper body, whereas whole-body motion, including facial expressions and finger movements should be integrated. Taking a long-term view of creating an artificial conversational agent, we would pursue integrating our model with other nonverbal behaviors and with a conversational model. Gestures are deeply related to verbalization according to the information packaging hypothesis \cite{kita2000representational}, so an integrated model generating speech and gestures together could deliver information more efficiently. 

\begin{acks}
The authors thank the anonymous reviewers for their thorough and valuable comments. This work was supported by the Institute of Information \& communications Technology Planning \& Evaluation (IITP) grant funded by the Korea government (MSIT) (No. 2017-0-00162, Development of Human-care Robot Technology for Aging Society). Resource supporting this work were provided by the `Ministry of Science and ICT' and NIPA (``HPC Support'' Project).
\end{acks}

\bibliographystyle{ACM-Reference-Format}
\bibliography{main}

\appendix

\begin{table}
\caption{The list of the gesture generation models used in the human evaluation. $\ast$ denotes the epoch having the best FGD.}
\label{tab:ut_models}
\centering
\begin{tabular}{p{0.54\linewidth}p{0.22\linewidth}S}
  \toprule
  Model & Training stage (epochs) & {FGD}\\ 
  \midrule
Proposed model & $89^\ast$ & 3.729\\
Proposed model without regularization terms & $83^\ast$ & 5.756\\
Proposed model without adversarial scheme & $87^\ast$ & 9.712\\
Proposed model without text modality & 20 & 12.144\\
Proposed model without audio modality & 16 & 16.558\\
Attentional Seq2Seq & $66^\ast$ & 18.054\\
Speech2Gesture & $86^\ast$ & 19.254\\
Joint embedding model & $98^\ast$ & 22.083\\
Proposed model without adversarial scheme and audio modality & 17 & 26.328\\
Attentional Seq2Seq & 20 & 28.273\\
  \bottomrule
\end{tabular}
\end{table}

\section{Detailed Architectures} \label{sec:detailedarch}

Figure \ref{fig:appendix_architecture} shows the detailed architectures of the encoders, gesture generator, and discriminator. Figure \ref{fig:appendix_architecture_autoencoder} shows the architecture of the feature extractor used in the Fr\'{e}chet gesture distance.

\section{Models in human evaluation} \label{sec:ut_models}
Table \ref{tab:ut_models} lists the models used in the human evaluation.

\afterpage{
\clearpage
\begin{figure}[p]
  \centering
  \includegraphics[trim=0 15 0 0,clip,width=0.87\linewidth]{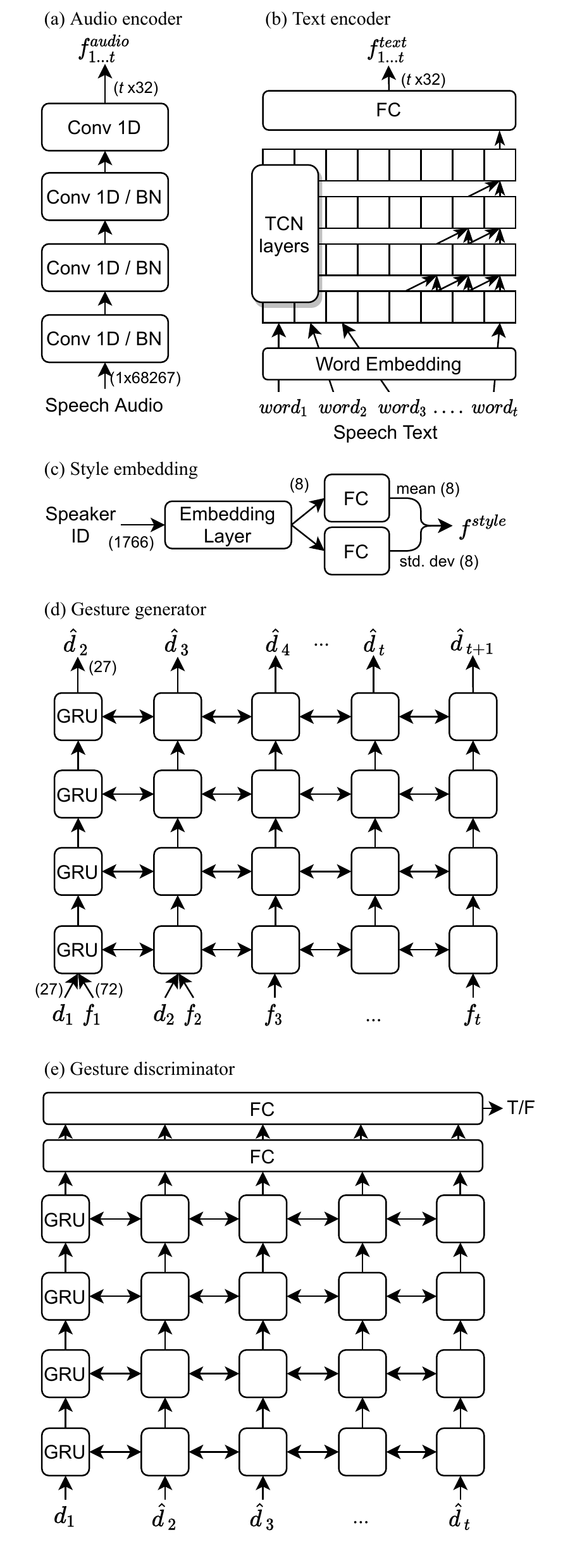}
  \caption{Detailed architectures of the (a) audio encoder, (b) text encoder, (c) speaker embedding, (d) gesture generator (assumed two seed poses), and (e) discriminator. BN stands for batch normalization, FC for fully connected layer, and TCN for temporal convolutional network.}
  \label{fig:appendix_architecture}
\end{figure}
\begin{figure}[p]
  \centering 
  \includegraphics[trim=0 60 50 0,clip,width=0.50\linewidth]{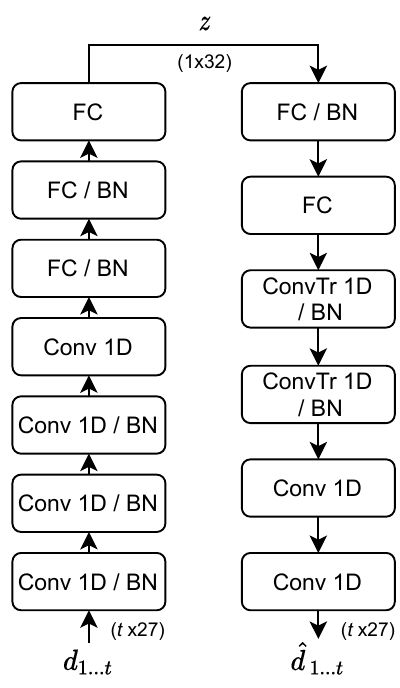}
  \caption{Detailed architecture of the autoencoder of the Fr\'{e}chet gesture distance. BN stands for batch normalization, FC for fully connected layer, and ConvTr for transposed convolution.}
  \label{fig:appendix_architecture_autoencoder}
\end{figure}
\clearpage
}

\end{document}